\documentclass[lettersize,journal]{IEEEtran}
\usepackage{amsmath,amsfonts}
\usepackage{array}
\usepackage{textcomp}
\usepackage{stfloats}
\usepackage{url}
\usepackage{verbatim}
\usepackage{graphicx}
\usepackage{cite}
\usepackage{authblk}
\usepackage{cite}
\usepackage{amsmath,amssymb,amsfonts}
\usepackage{graphicx}
\usepackage{textcomp}
\usepackage{xcolor}
\usepackage{tabularx}
\usepackage{url}
\usepackage[bookmarks=false]{hyperref}
\usepackage{rotating}
\usepackage{subcaption}
\usepackage{multirow}
\usepackage{pifont}
\usepackage{tcolorbox}
\definecolor{tumColorLightBlue}{HTML}{f0f5fa} 
\usepackage{cleveref}
\usepackage{algorithm}
\usepackage{algpseudocode}

\usepackage{tablefootnote}
\usepackage{threeparttable}
\usepackage{multirow}
\usepackage{booktabs}
\usepackage{harveyballs}
\usepackage{fancyhdr}

\begin{document}

\title{Towards an Optimized Multi-Cyclic Queuing and Forwarding in Time Sensitive Networking with Time Injection}

\author{Rubi Debnath,~\IEEEmembership{Graduate Student Member,~IEEE,} Mohammadreza Barzegaran,~\IEEEmembership{Member,~IEEE,} Sebastian Steinhorst,~\IEEEmembership{Senior Member,~IEEE,}}



\pagestyle{fancy}
\fancyhf{}  
\renewcommand{\headrulewidth}{0pt}

\fancypagestyle{firstpage}{
\fancyhf{}  
\renewcommand{\headrulewidth}{0pt}
\fancyhead[C]{\small
This paper has been submitted to IEEE for possible publication. Copyright may be transferred without notice, after which this version may no longer be accessible. Permission from IEEE must be obtained for all uses.}
}

\maketitle
\thispagestyle{firstpage} 
\pagestyle{empty}

\begin{abstract}
Cyclic Queuing and Forwarding (CQF) is a Time-Sensitive Networking (TSN) shaping mechanism that provides bounded latency and deterministic Quality of Service (QoS). However, CQF's use of a single cycle restricts its ability to support TSN traffic with diverse timing requirements. Multi-Cyclic Queuing and Forwarding (Multi-CQF) is a new and emerging TSN shaping mechanism that uses multiple cycles on the same egress port, allowing it to accommodate TSN flows with varied timing requirements more effectively than CQF. Despite its potential, current Multi-CQF configuration studies are limited, leading to a lack of comprehensive research, poor understanding of the mechanism, and limited adoption of Multi-CQF in practical applications. Previous work has shown the impact of Time Injection (TI), defined as the start time of Time-Triggered (TT) flows at the source node, on CQF queue resource utilization. However, the impact of TI has not yet been explored in the context of Multi-CQF. This paper introduces a set of constraints and leverages Domain Specific Knowledge (DSK) to reduce the search space for Multi-CQF configuration. Building on this foundation, we develop an open-source Genetic Algorithm (GA) and a hybrid GA-Simulated Annealing (GASA) approach to efficiently configure Multi-CQF networks and introduce TI in Multi-CQF to enhance schedulability. Experimental results show that our proposed algorithms significantly increase the number of scheduled TT flows compared to the baseline Simulated Annealing (SA) model, improving scheduling by an average of 15\%. Additionally, GASA achieves a 20\% faster convergence rate and lower time complexity, outperforming the SA model in speed, and efficiency.
\end{abstract}


\section{Introduction}
\label{sec:introduction}
\begin{figure}[!t]
    \centering
    \includegraphics[scale=0.23, trim={0.5cm 0.5cm 0.5cm 0.5cm}, clip]{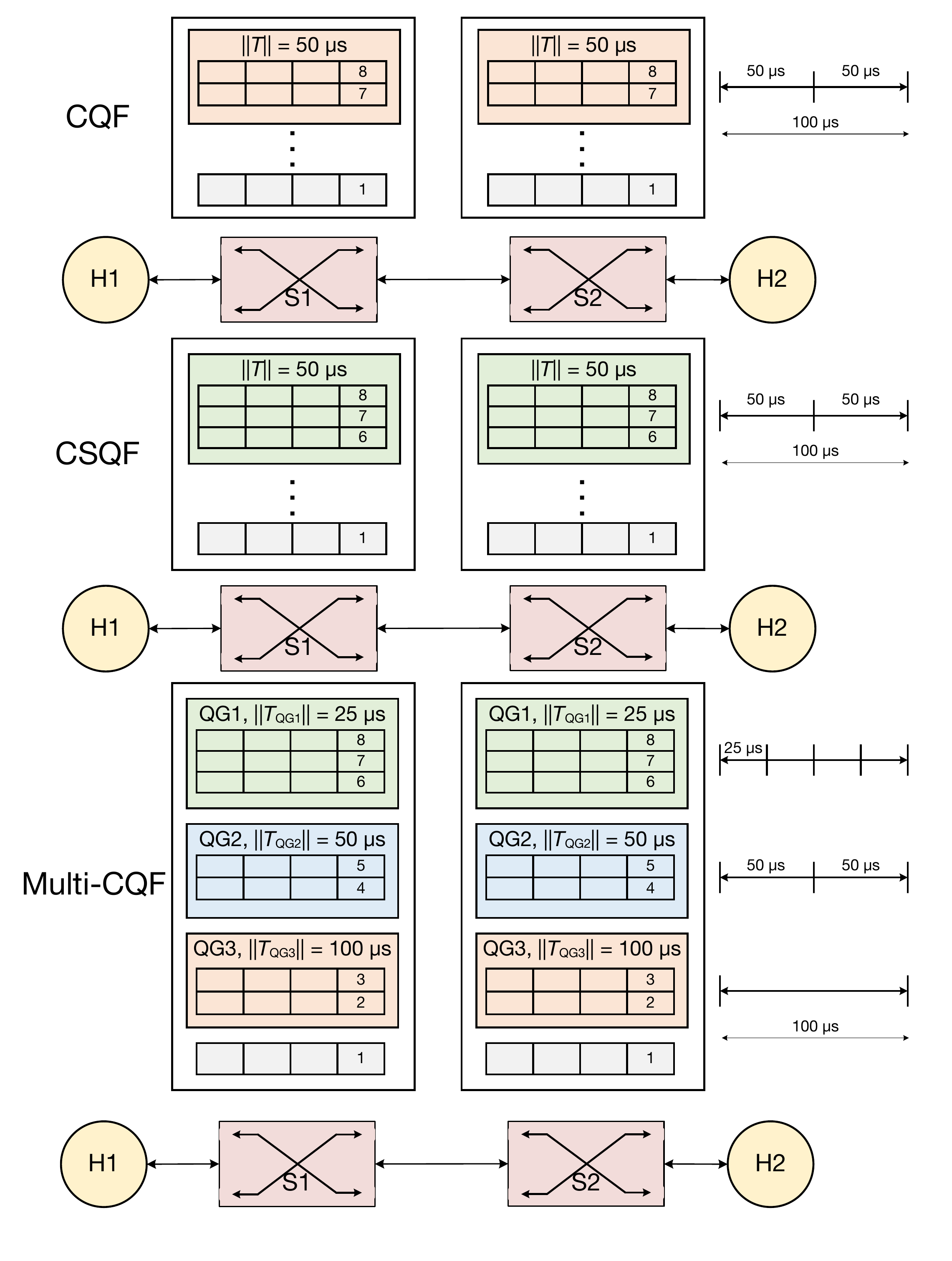}
    \caption{A network with two switches (S1 and S2) and a source (H1) and destination node (H2), showcasing the CQF, CSQF and the Multi-CQF architecture with different cycle and queue.}
    \label{fig:tsn_mechanism}
    \vspace{-0.3cm}
\end{figure}

\IEEEPARstart{W}{ith} the rapid advancement of communication technologies and the growing demand for safety-critical applications, modern communication systems must support multiple traffic types with diverse Quality of Service (QoS) requirements in the same network. Time-Sensitive Networking (TSN)~\cite{8021Q} has become popular for accommodating mixed-criticality traffic~\cite{rubi_icc} through various shaping and scheduling mechanisms on the same egress port, making it ideal for industrial, vehicular, 5G-TSN~\cite{rubi_vtc}, aerospace, and other high-demand safety-critical networks. In a TSN network, there are eight queues in each egress port~\cite{8021Q} as shown in Fig.~\ref{fig:tsn_mechanism}. Depending on the shaping and scheduling mechanism in the TSN network, the architecture and the functionality changes (refer Fig.~\ref{fig:tsn_mechanism}). 
Time Aware Shaper (TAS)~\cite{8021Qbv, rubi_rtcsa} is a popular TSN scheduling mechanism, working on the principle of the static table, known as the Gate Control List (GCL), which provides bounded latency and low jitter. However, the generation of the GCL for a TAS enabled TSN network is highly complex and computationally expensive. 

\begin{figure*}[!t]
    \centering
    \begin{minipage}[b]{0.32\textwidth}
    \centering
    \includegraphics[scale=0.18, trim={1cm 0cm 1cm 6mm}, clip]{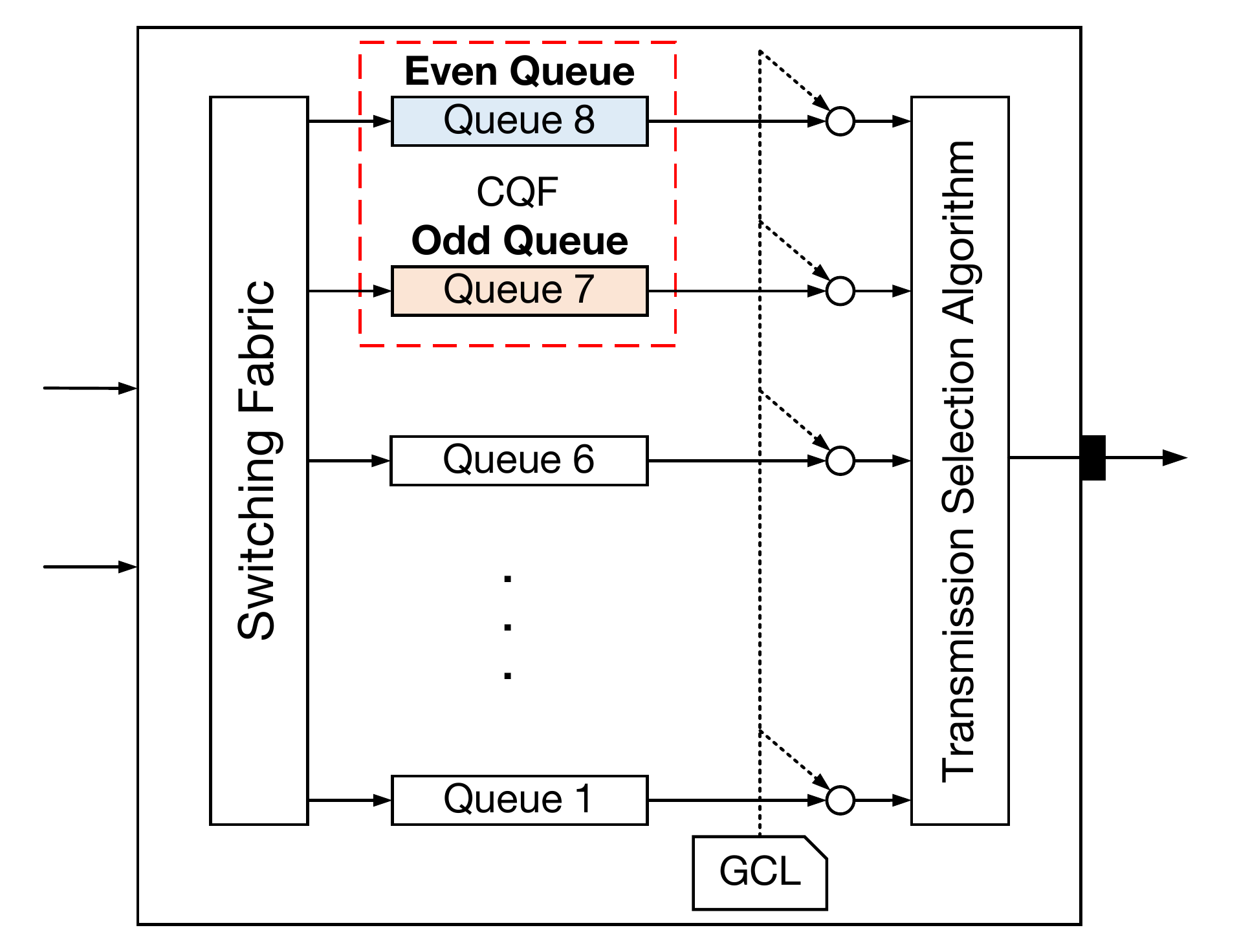}
    \subcaption{CQF Architecture}
    \label{fig:cqf_overview}
    \end{minipage}
    \begin{minipage}[b]{0.32\textwidth}
    \centering
    \includegraphics[scale=0.18, trim={1cm 0cm 1cm 7mm}, clip]{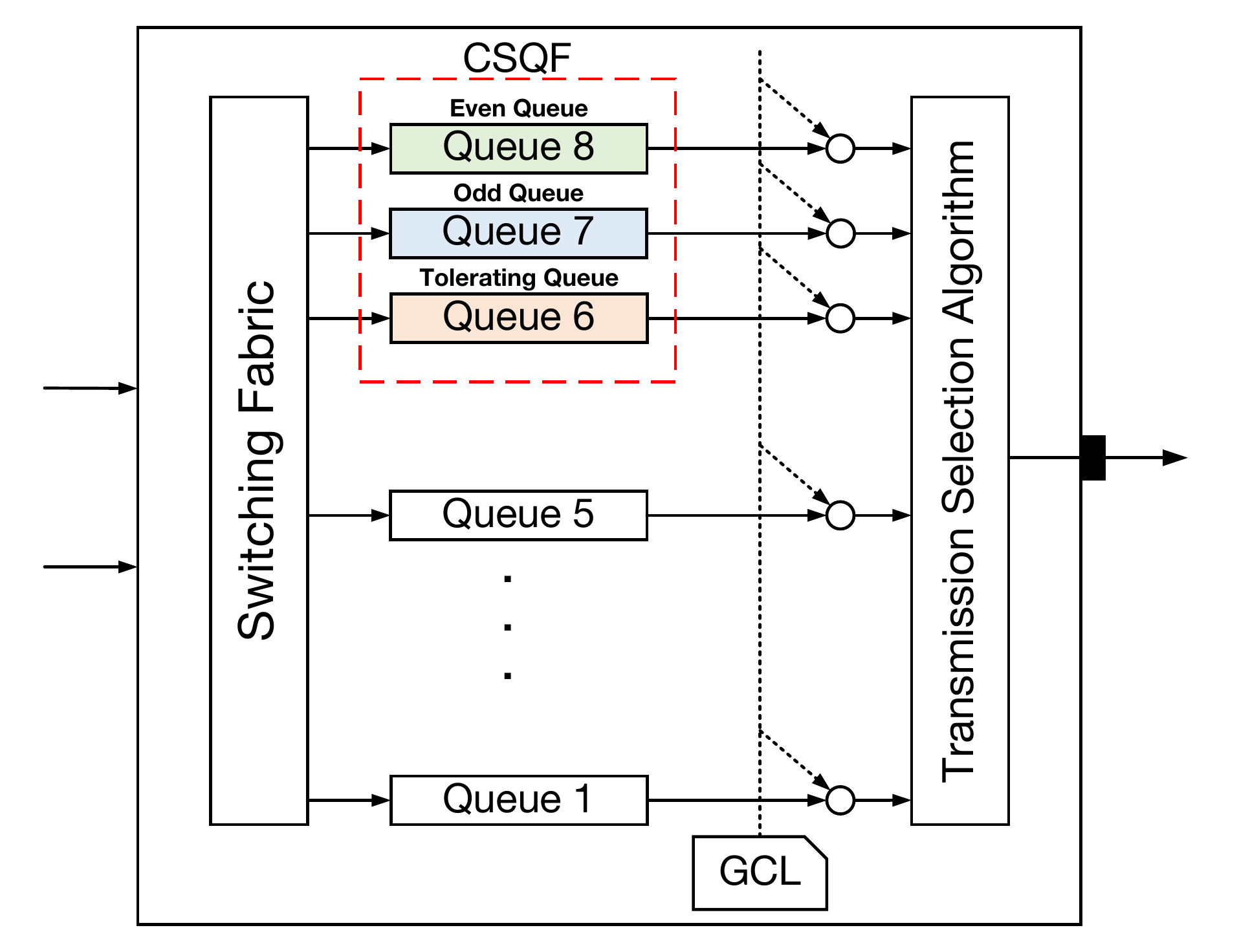}
    \subcaption{CSQF Architecture}
    \label{fig:csqf_overview}
    \end{minipage}
    \begin{minipage}[b]{0.32\textwidth}
    \centering
    \includegraphics[scale=0.18, trim={1cm 0cm 1cm 7mm}, clip]{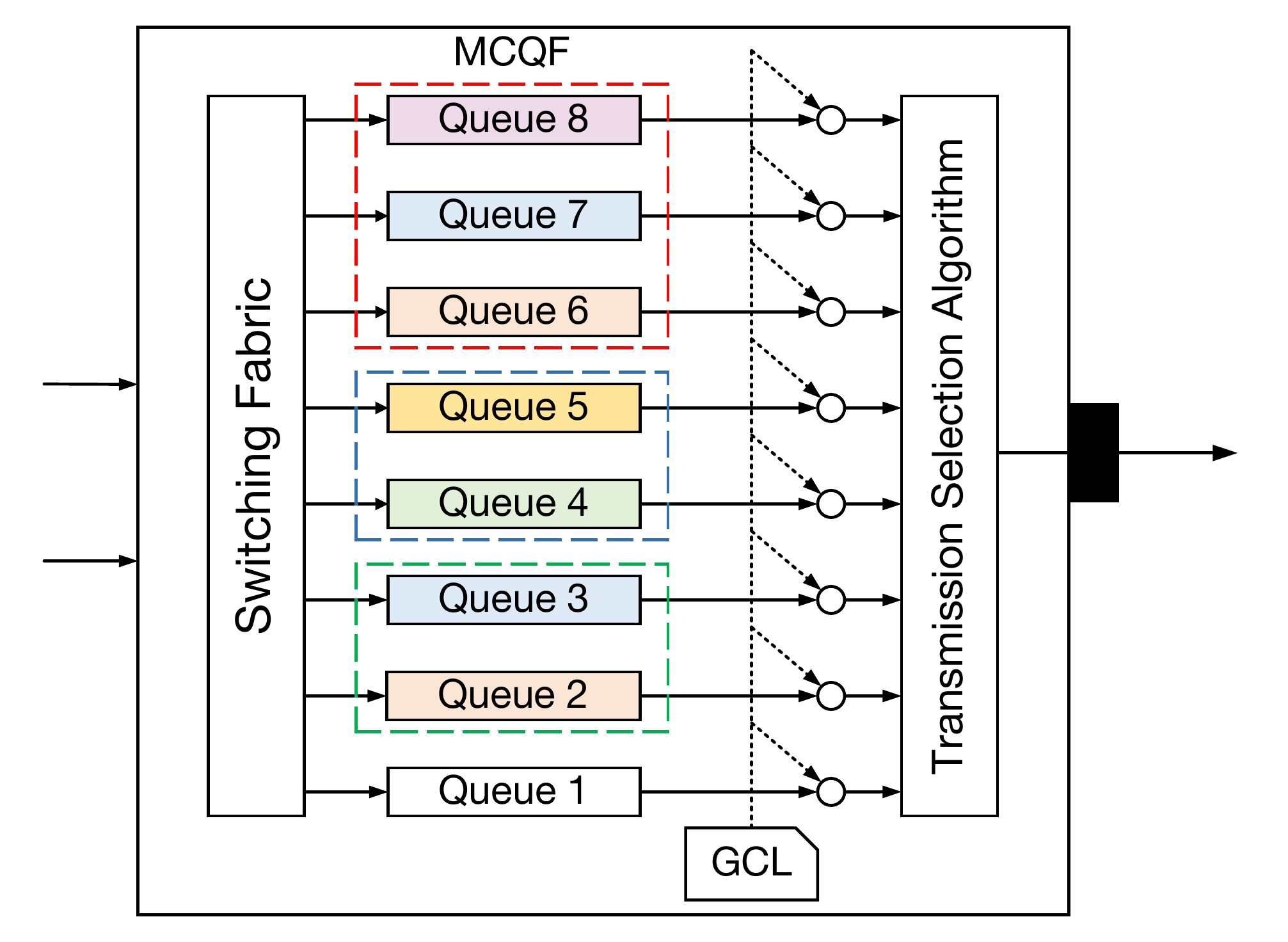}
    \subcaption{Multi-CQF Architecture}
    \label{fig:mcqf_overview}
    \end{minipage}
    \caption{Overview of the different cyclic shaper variants.}
    \vspace{-0.3cm}
\end{figure*}

Cyclic Queuing and Forwarding (CQF)~\cite{8021Qch}, another TSN shaping mechanism, emerged as an alternative to TAS, operating cyclically with two queues denoted as \textit{even} and \textit{odd} queue. Like TAS, CQF also relies on a GCL, however, unlike TAS, where gate opening and closing times vary, CQF follows a fixed open and close time. Therefore, for CQF, we do not calculate the GCL, as the gates are configured based on the cycle. Additionally, for CQF, we use two queues for the same priority and traffic type~\cite{boyer_2024_cqf_cycle}. The primary advantage of CQF lies in its simple and static gate operations (\texttt{open} and \texttt{close}) in a ping-pong manner, providing bounded and deterministic delay. This simplicity sets CQF apart from TAS, which, while offering stricter delay and jitter control, involves significantly higher complexity in generating the GCL, especially as the number of Time-Triggered (TT) flows and network size increase. As shown in Fig.~\ref{fig:tsn_mechanism} and \ref{fig:cqf_overview}, CQF operates with a single cycle and two queues (one receiving and one transmitting at any given cycle). The end-to-end delay of TT flows in a CQF network depends on the number of hops in the route and the cycle. Therefore, the cycle has a direct impact on the performance of the CQF network. Despite CQF's ease of operation, selecting an appropriate cycle in CQF is critical: a smaller cycle may not support a high volume of flows arriving from different senders and going to the same egress port, while a larger cycle accommodates more flows but increases the Worst-Case Delay (WCD). The WCD is defined as the maximum end-to-end delay of the flows from the source to the destination. 

Cycle Specific Queuing and Forwarding (CSQF)~\cite{ietfSegmentRouting, rubi_ccnc} is another cyclic shaper mechanism which overcomes the challenges of time synchronization error, and long and short propagation delay in Wide Area Networks (WANs). In local area TSN networks, the nodes and the switches are strictly synchronized using a time synchronization mechanism~\cite{8021AS}. However, in WAN, unlike TSN, the network cannot be strictly time synchronized. Furthermore, in the network, there are long and short Ethernet cables leading to large and small propagation delays respectively, making it challenging to predict the mentioned delays in advance during the cycle selection. This difference in propagation delay and time synchronization error results in flows arriving to the nodes out of their designated cycle. To overcome this challenge, CSQF introduces a third queue, known as the \emph{tolerating queue}, for the flows which arrive outside of their cycle. CSQF uses multiple receiving queues to compensate for the additional delay caused by long-distance links~\cite{csqf1}. Different from CQF, in CSQF, we use three queues (see Fig.~\ref{fig:csqf_overview}) for the same traffic type and priority. During each cycle, two queues receive incoming flows, while the third transmits flows from the previous cycle. \textit{Despite the addition of a third queue, CSQF still operates with a single cycle, which limits its ability to support diverse timing requirements for different traffic types, similar to CQF.} Since real-world cases are based on mixed-criticality networks, relying on a single cycle limits the practical application of the cyclic shapers, highlighting the need for multi-cycle adoption.

Multi-Cyclic Queuing and Forwarding (Multi-CQF)~\cite{norman}, depicted in Fig.~\ref{fig:mcqf_overview}, is an emerging cyclic shaping mechanism that operates with multiple queue groups and cycles. Each queue group can be an instance of CQF or CSQF and works with its own unique cycle. Hence, Multi-CQF can be viewed as a collection of multiple CQF and/or CSQF instances coexisting on the same egress port. As shown in Fig.~\ref{fig:mcqf_overview}, the illustrated Multi-CQF architecture includes one CSQF and two CQF instances. However, this architecture is not strict and can be configured with any combination of CQF and CSQF. Multi-CQF overcomes the limitations of other cyclic shapers by supporting traffic with diverse timing requirements through the use of multiple instances of CQF and CSQF operating on different cycles. Unlike CQF and CSQF, Multi-CQF can use up to seven queues for the same traffic type and priority (refer to Fig.~\ref{fig:mcqf_overview}), although other combinations are possible. 

Alexandris et al. in \cite{mcqf_paul} introduced algorithms based on Constraint Programming (CP) and Simulated Annealing (SA) for Multi-CQF configuration. But, \cite{mcqf_paul} did not address some critical aspects such as: (1) the constraints for determining cycle, (2) the impact of Time Injection (TI) (TI is defined as the start time or offset of the TT flow in the source node) on Multi-CQF, (3) the impact of flow sorting, flow-to-queue-group assignment, and other domain-specific knowledge (DSK) of Multi-CQF. Notably, \cite{mcqf_paul} assumes the cycle to be predefined. In addition, the utilization of TI to maximize resources in Multi-CQF remains an open research question. As shown in~\cite{jrs_cqf_transaction}, the scheduling and configuration of cyclic shapers are heavily influenced by the order of the flow sorting. Furthermore, assigning flows to appropriate queue groups introduces further complexity in configuring Multi-CQF. By leveraging DSK, flows can be intelligently distributed into queue groups, which significantly reduces the configuration search space and enables faster algorithmic convergence. We discuss this in detail in Section~\ref{sub:ss}.

In this paper, we address the gaps by defining and formulating the necessary constraints for an optimized configuration of Multi-CQF-enabled networks. We first determine the cycle combination by incorporating our proposed constraints and then present a DSK-guided GA and a hybrid GASA algorithm to configure Multi-CQF networks. Using the appropriate cycle increases the overall schedulability of the flows and further using the DSK reduces the search space, thereby, reducing the computation time resulting in faster convergence of the optimization function. Additionally, we introduce TI to increase schedulability and evaluate the performance of Multi-CQF. Our proposed algorithms outperform the baseline SA of \cite{mcqf_paul} by scheduling more TT flows, achieved through the use of appropriate cycle selection and flow-to-queue-group assignment. The sorting of flows into different queue groups and the use of the DSK reduces the search space, leading to faster convergence of the objective function. The proposed models are adaptable to various topologies and timing requirements. In particular, our main contributions are: 

\begin{enumerate}
    \item \textbf{Constraints and DSK:} We introduce and formulate constraints and utilize DSK to optimize the Multi-CQF search space, improving the configuration solution and time complexity (Section~\ref{sec:problem}, \ref{sub:flow_to_qg_mapping} and \ref{sub:ss}). 
    \item \textbf{Efficient Cycle Combinations:} We identify efficient cycle combinations using a constraint-guided heuristic search approach to increase the overall schedulability (Section~\ref{sec:cycle_model}).
    \item \textbf{Time Injection (TI):} We implement TI for Multi-CQF and evaluate its impact on schedulability and configuration (Section~\ref{sec:problem} and \ref{sec:implementation}). 
    \item \textbf{Configuration Algorithm:} We propose a publicly accessible open source\footnote{Github:\url{https://github.com/tum-esi/gasa-mcqf-ti}} metaheuristic-based GA and a hybrid GA+SA model called GASA for Multi-CQF configuration (Section~\ref{sec:ga} and \ref{sec:gasa}).
    \item \textbf{Extensive Evaluation:} We implement the SA model from \cite{mcqf_paul} as a baseline for comparison and conduct extensive experiments to validate the performance of our proposed models (Section~\ref{sec:results}).
\end{enumerate}

\section{Related Work}
\label{sec:related_work}
Related work in \cite{rubi_rtcsa, luxi_tnsm, rubi_noms} evaluated the different shapers and schedulers in TSN both individually and in combination. They further demonstrated the benefits of using joint shapers, emphasizing the need to evaluate all shapers in an egress port. However, these studies did not cover CQF, CSQF, or Multi-CQF individually or in combination. Nasrallah et al. conducted a comprehensive survey on cyclic queuing mechanisms for large-scale deterministic networks (LDNs) in \cite{survey_cqf_ahmed}. This survey covers work on cyclic shapers up to 2018, however, most papers on CQF and other cyclic shapers were published after 2019. Therefore, the survey is outdated and does not include many recent works on cyclic shapers in TSN.

Scheduling and configuration in TSN are optimization problems, often solved using mathematical solvers such as satisfiability modulo theories (SMT). However, mathematical solvers, and Integer Linear Programming (ILP) methods have limitations in solving large scale scheduling problems in TSN~\cite{ilp_limitations}. As a result, heuristic and metaheuristic methods have emerged as popular alternatives, offering near-optimal solutions with reasonable time complexity. Among these methods, Genetic Algorithms (GAs) and Simulated Annealing (SA) are used to solve complex computational models and generate optimal or near-optimal solutions. Previous studies \cite{ga_tt_pahlevan, ga_anna} have successfully applied GAs to solve the TAS scheduling problem in TSN. However, for cyclic shapers GAs are yet to be explored.

Several works have explored scheduling enhancements and flow optimization in CQF networks. Yan et al. in \cite{itp_infocom} proposed an Injection Time Planning (ITP) algorithm to improve the schedulability of TT flows in CQF networks, demonstrating that the start time of the flows significantly impacts the utilization of CQF queue resources. They introduced TI, where flows are sent at specific time offsets from the source node. However, Multi-CQF and the impact of TI on Multi-CQF has not been explored in \cite{itp_infocom}. In Multi-CQF, TI is dependent on the specific cycle of each queue group, making it queue group-specific. Hence, despite the study of TI in CQF, the study of Multi-CQF is still required.

Quan et al. in \cite{online_qch_quan} proposed a scheduling algorithm called \textquotedblleft Fits\textquotedblright\ to incrementally schedule TT flows in CQF networks. Nevertheless, the existing CQF solutions will not work with Multi-CQF due to the use of multiple cycle and distinct queue groups with multiple instances of CQF and/or CSQF. Wang et al. in \cite{jrs_cqf_transaction} jointly optimized the routing and scheduling of CQF networks. The sorting of TT flows significantly impacts the overall schedulability of a CQF network. In \cite{jrs_cqf_transaction}, the authors proposed a normalization method to calculate the weight of TT flows based on their parameters such as deadline, periodicity, and payload size. They then sorted the TT flows according to these weights and introduced a load-balancing algorithm to improve the scheduling success ratio and optimize network resource utilization. While the sorted TT flows are configured in the CQF network in \cite{jrs_cqf_transaction}, Multi-CQF consists of multiple CQF or CSQF instances, making this approach unsuitable for Multi-CQF. Multi-CQF also requires the flow-to-queue-group mapping other than the flow sorting. Hence, flow sorting and flow-to-queue-group mapping has to be studied separately for Multi-CQF.

Multiple studies have aimed to study the importance and the impact of cycle in cyclic shapers. Guidolin-Pina et al. in \cite{boyer_2024_cqf_cycle} provided the mathematical proof and the condition for the minimum cycle in CQF. They further showed that CQF can absorb burstiness of the flows as long as the cycle is large enough and the link capacity or bandwidth (BW) is available. This highlights the importance of proper cycle selection in CQF to ensure the correct design and effective operation of cyclic shapers. They further discussed in \cite{boyer_cqf_guard_band} that the burstiness of flows in a cycle does not increase as the flows progress through the network. Huang et al. in \cite{cqf_challenges} further combined layer-2 CQF and the layer-3 CSQF in the same network. They proposed a novel architecture C-TSDN, in which the CQF and the CSQF are jointly designed to solve the end-to-end scheduling. They further discussed in \cite{drl_mcqf} the issue of resource wastage in CSQF due to the use of the same cycle for both high and low speed links. In high speed links, a smaller cycle is sufficient, whereas in low speed links, a larger cycle is required. Therefore, to solve this problem they proposed multi-cycle CSQF (MCCSQF) mechanism for multi-link rate networks suggesting to use different cycles for different links with different BW. They designed a DRL framework for routing and scheduling in MCCSQF. However, contrary to MCCSQF, Multi-CQF in this paper uses different cycles with multiple instances of CQF and/or CSQF on all links irrespective of its link BW. Together, these related works highlight the fundamental impact of cycle configuration on the performance, scalability, and efficiency of cyclic shapers like CQF and CSQF. Furthermore, it is evident from the related work that one cycle is not sufficient for different network scenarios.

In addition to earlier studies, recent research has increasingly explored modified variants of CQF. Yang et al. in \cite{burst_aware_mcqf} aimed to schedule both TT and bursty flows in a CQF network. As CQF cannot accommodate bursty flows, they proposed an enhanced Multi-CQF model which supports both bursty and TT traffic. The enhanced Multi-CQF model has three or more than three queues on every switch port and one cycle, and most importantly, it only has one instance of three or more than three queue CQF. During each cycle, one queue transmits and all the other queues receive the flows. In contrast, we use different cycle and queue groups in our Multi-CQF network. It is important to note that the proposed MCCSQF in \cite{drl_mcqf} and the enhanced Multi-CQF in \cite{burst_aware_mcqf} are different from the Multi-CQF discussed in this paper. Most importantly, there is a clear need to standardize the nomenclature of these mechanisms to avoid confusion across studies and implementations.

Konstantinos et al. in \cite{mcqf_paul} proposed a CP and SA-based metaheuristic solution for CQF, CSQF, and Multi-CQF, conducting a comprehensive evaluation of the different cyclic shaper variants. Although they introduced an SA-based Multi-CQF configuration model, their study does not address several key aspects such as cycle selection, flow-to-queue-group assignment, and TI. Building on \cite{mcqf_paul}, Pop et al. in \cite{paul_mcqf_combination} formulated the combinatorial optimization problem of configuring TAS with Multi-CQF in the network. In our prior work~\cite{rubi_ccnc}, we conducted the first OMNeT++ based simulation using our open-source framework to evaluate the performance of all cyclic shaper variants: CQF, CSQF, and Multi-CQF. Our simulation results showed that CSQF schedules a larger number of TT flows at the cost of increased end-to-end delay in the network, and Multi-CQF clearly supports traffic type with diverse timing requirements. These findings, together with the growing interest in cyclic shaping mechanisms, underline the need for a deeper and more systematic evaluation of Multi-CQF configuration.

\begin{figure}
    \centering
    \includegraphics[scale=0.24, trim={0cm 3cm 1.7cm 3.2cm}, clip]{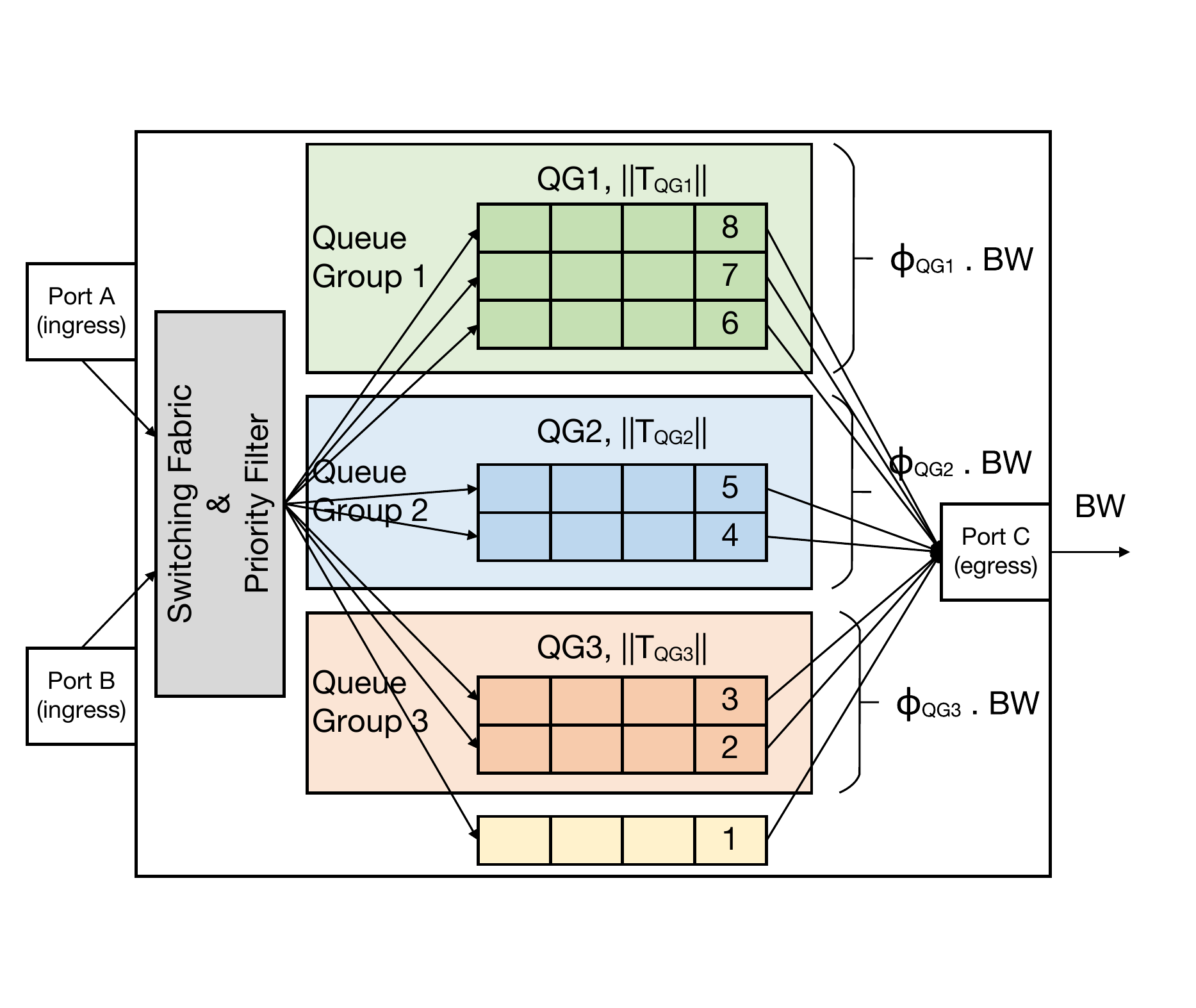}
    \caption{Multi-CQF Framework with different queue groups.}
    \label{fig:mcqf_bw}
    \vspace{-0.3cm}
\end{figure}

\section{System Model}
\label{sec:model} 
Our TSN network consists of switches \texttt{(SWs)} and end stations \texttt{(ESs)}. Together the \texttt{(SWs)} and \texttt{(ESs)} are referred to as nodes. We represent the TSN network as an undirected graph denoted with $\mathcal{G(V, E)}$ where $\mathcal{V}$ denotes the nodes or vertices and $\mathcal{E}$ denotes the edges or links. $\mathcal{V}$ consists of \texttt{(ESs)} and \texttt{(SWs)} given as $\mathcal{V}$ = \texttt{(ESs $\cup$ SWs)} and $\mathcal{E}$ consists of all the links between the nodes in the network given as $e_j$ $\in$ $\mathcal{E}$ where $j$ denotes the link/edge number. This paper considers periodic flows with hard real-time requirements and assumes that these flows are scheduled using cyclic shapers, where all flows have the highest priority. Each flow sends repeated data based on its periodicity known as the frame. $\mathcal{F}$ is the set of all TT flows, where a total of $\mathcal{|F|}$ flows are present in the network. Each TT flow $(f_{i})$ is defined as a tuple of: $\forall f_{i} \in$ $\mathcal{F}$, \text{where} $i = 1 \ldots$ $\mathcal{|F|}$
\begin{equation}
    f_{i} = \langle id, src, dst, period, deadline, size \rangle,
\end{equation}
\noindent where $id$ is the flow number, $src$ is the source, $dst$ is the destination, $period$ is the periodicity of the flow in $\mu$s, $deadline$ is the deadline in $\mu$s, and $size$ is the frame length of the flow in Bytes. We consider the  size consists of the frame size of the flows (payload and the headers in Bytes). In our model, we assume that the network is time-synchronized using gPTP~\cite{8021AS}. However, to compensate for the time synchronization error, we have considered the synchronization error and other delays in the network. Each flow $f_i$ is sent from the $src$ to the $dst$ using the route $(r_s \in R_{f_i})$ and $R_{f_i} \subseteq R$. The route $R_{f_i}$ consists of \texttt{K-shortest paths} using Yen's algorithm~\cite{yen} for flow $f_i$ from the $src$ to the $dst$ and $R$ is the set of all routes for all TT flows. 

\subsection{Multi-Cyclic Queuing and Forwarding (Multi-CQF)}
\label{sub:mcqf} 
Multi-CQF is a new variant of cyclic shaper, and Fig.~\ref{fig:mcqf_bw} illustrates its working mechanism and architecture as used in this paper. Multi-CQF consists of different Queue Groups, denoted as (\texttt{QG}s) in this paper, as shown in Fig.~\ref{fig:mcqf_overview} and Fig.~\ref{fig:mcqf_bw}. \texttt{QG} is defined as a group consisting of two or more than two queues operating with one cycle ($T_{\texttt{QG}x}$), where $x$ represents the \texttt{QG} number ($x \in 1,2,3$). Multi-CQF consists of one or more instances of CQF and/or CSQF, where each instance is called a \texttt{QG}. All the \texttt{QG}s together should not send flows more than the BW of the network. Therefore, to prevent the over-subscription of the BW, each \texttt{QG} is allocated a configurable percentage of BW (see Fig.~\ref{fig:mcqf_bw}). Furthermore, each \texttt{QG} transmits flows within its assigned BW limits. In the presented Multi-CQF architecture, Queue 0 is unused and left for other traffic types, such as Best Effort (BE) (refer to Fig.~\ref{fig:mcqf_bw}). In our work, we consider TT flows which are periodic and non-bursty traffic types in nature. Depending on the architecture and the priorities considered in the network, Multi-CQF flows may experience blocking from either higher or lower priority traffic. However, as long as the \texttt{QG}s are sending flows within their BW limits, all the flows will be de-queued within their cycle window, thereby not causing any interference from the lower and the higher priority flows~\cite{norman}. Multi-CQF involves several critical design choices, including (1) cycle selection, (2) flow sorting and flow-to-\texttt{QG} mapping, and (3) BW distribution among the \texttt{QG}s. Each of these design aspects represents a separate optimization problem, and solving all of them simultaneously within a single framework becomes highly complex. Therefore, in this paper, we treat some of these design choices as fixed input parameters to reduce complexity and focus on scheduling and cycle selection. We would also like to emphasize that determining the minimum or optimal Multi-CQF cycle ($T_{\texttt{QG}x}$) is a separate mathematical optimization problem and is beyond the scope of this paper. Instead, our approach identifies an efficient or near-optimal cycle combination ($T_{\texttt{QG}x}$) using a set of well-defined constraints and DSK-guided heuristic search. We discuss these design choices in detail in Section~\ref{sec:problem}. We now present the worst-case queuing delay and worst-case delay of Multi-CQF as follows: 

\begin{figure}[t!]
    \centering
        \includegraphics[scale=0.2, trim={0.5cm 0.5cm 0.5cm 0.5cm}, clip]{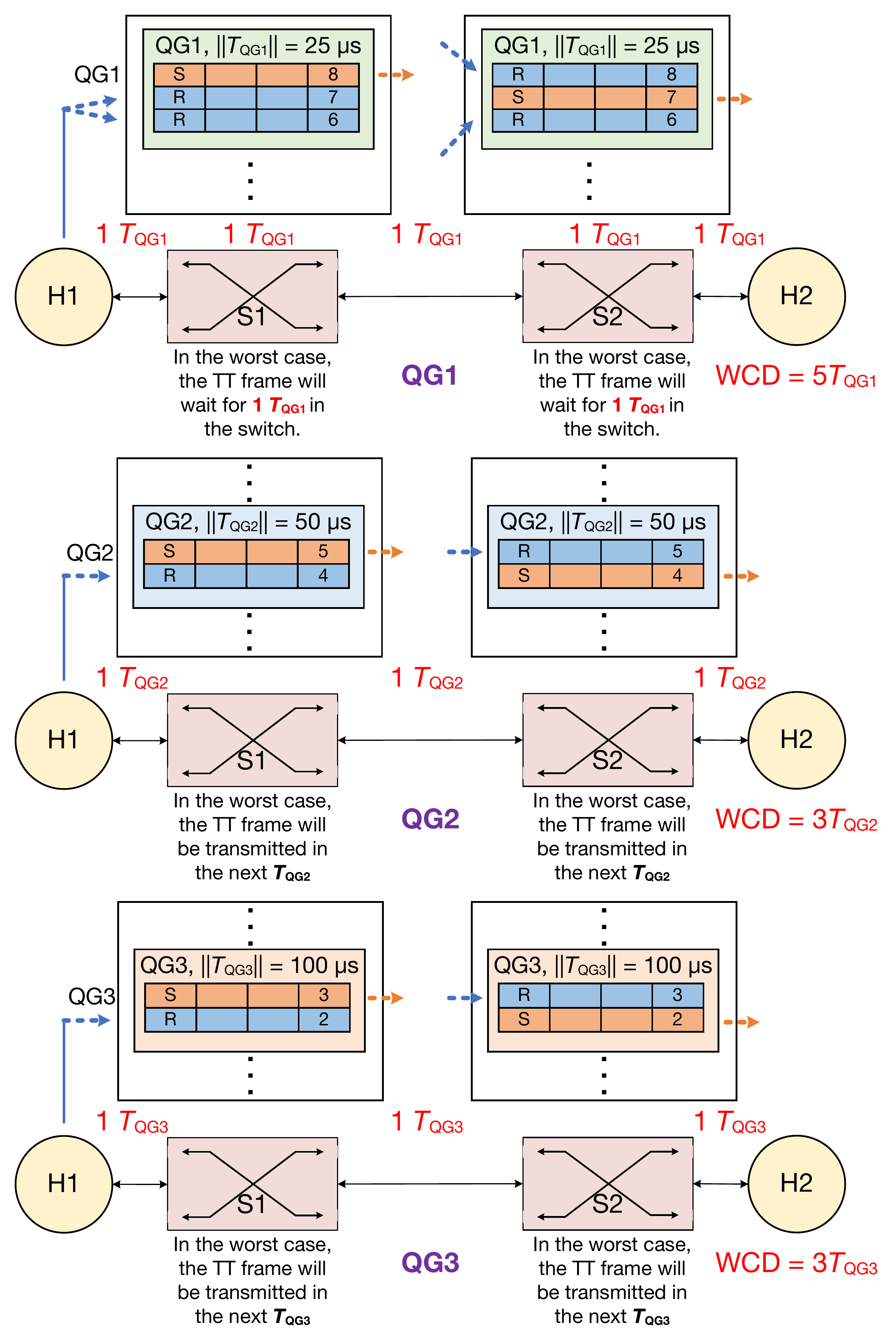}
        \caption{Worst-Case Queuing Delay for \texttt{QG}1, \texttt{QG}2, and \texttt{QG}3 in Multi-CQF. In \texttt{QG}1, one queue transmits (S) while two receive (R). In the worst case, flows in \texttt{QG}1 are stored in the \emph{tolerating queue} at each switch, causing an extra cycle of delay before transmission. In contrast, flows in \texttt{QG}2 and \texttt{QG}3 are forwarded in the next cycle.}
    \label{fig:mcqf_all_qg_wcd}
    \vspace{-0.3cm}
\end{figure}

\vspace{0.5em}
\noindent \textbf{Worst-Case Queuing Delay ($\mathrm{d_{queue}^i}$):} The frames experience queuing delay in the switch along the route. This delay represents the total amount of time the TT flow frames spend waiting in the egress port queue before transmission. The maximum queuing delay of a TT flow $f_i$, denoted as $\mathrm{d_{queue}^i}$, depends on the \texttt{QG} of the flow and its cycle. Therefore, we express $\mathrm{d_{queue}^i}$ in terms of the cycle as follows:
\vspace{-0.1cm}
\begin{equation}
  \mathrm{d_{queue}^i} =
    \begin{cases}
      T_{\texttt{QG}x} & , \forall \; f_i \in \texttt{QG}x, \; \text{if \texttt{QG}x has 3 Queues,} \\
      0  & , \; \mathrm{otherwise}. \\
    \end{cases}
    \label{eq:dqueue}
\end{equation}
\noindent where $x = 1,2,3$ for TT flows belonging to \texttt{QG}1, \texttt{QG}2, and \texttt{QG}3 respectively.

\begin{figure}[t!]
    \centering
        \includegraphics[scale=0.23, trim={0.5cm 0.5cm 0.5cm 0.1cm}, clip]{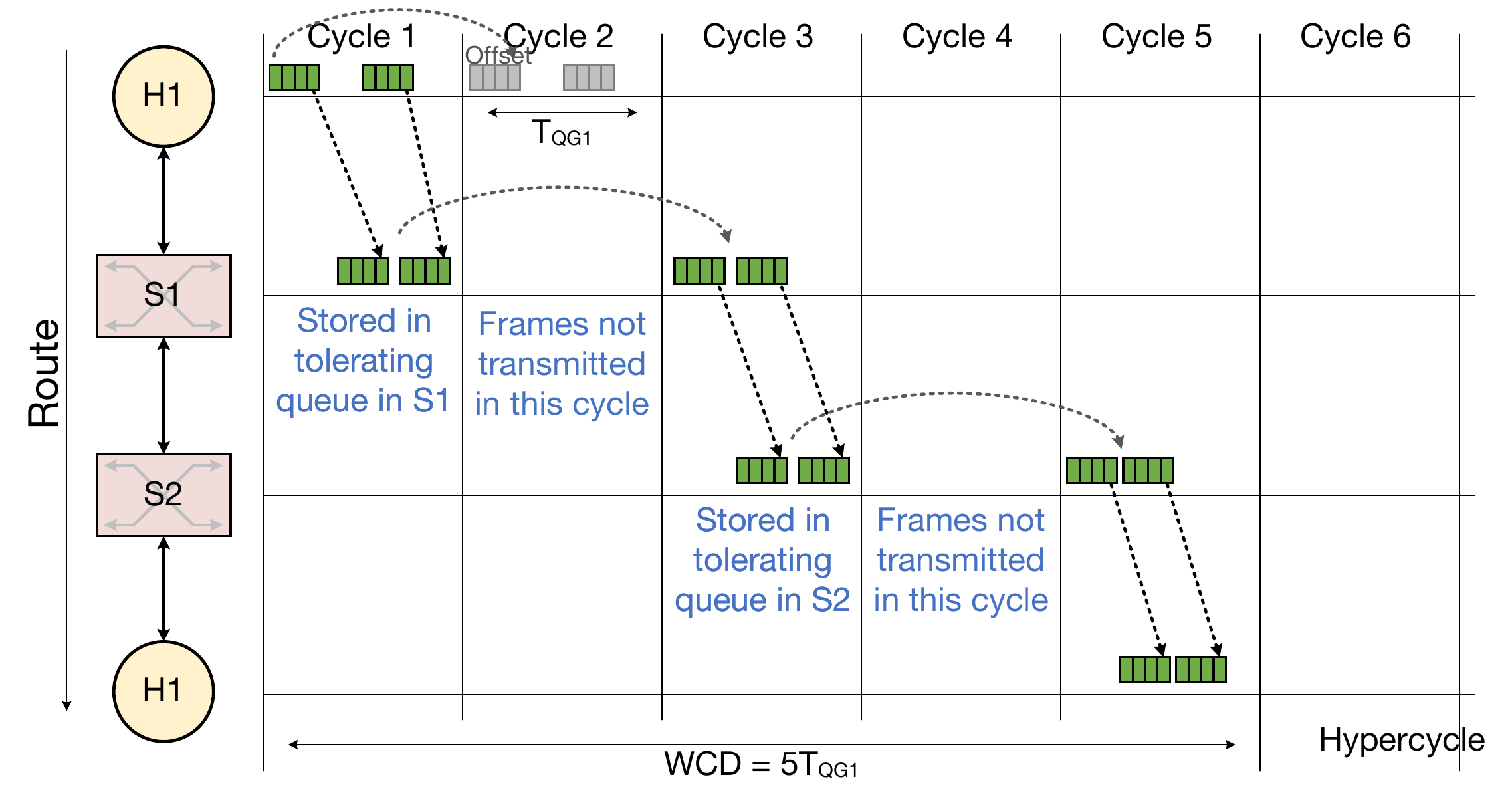}
        \caption{An illustration of flow transmission in a holistic scenario for \texttt{QG}1. The green boxes represent the flows.}
    \label{fig:qg1_wcd}
    \vspace{-0.2cm}
\end{figure}

\begin{figure}[t!]
    \centering
        \includegraphics[scale=0.23, trim={0.5cm 0.5cm 0.5cm 0.5cm}, clip]{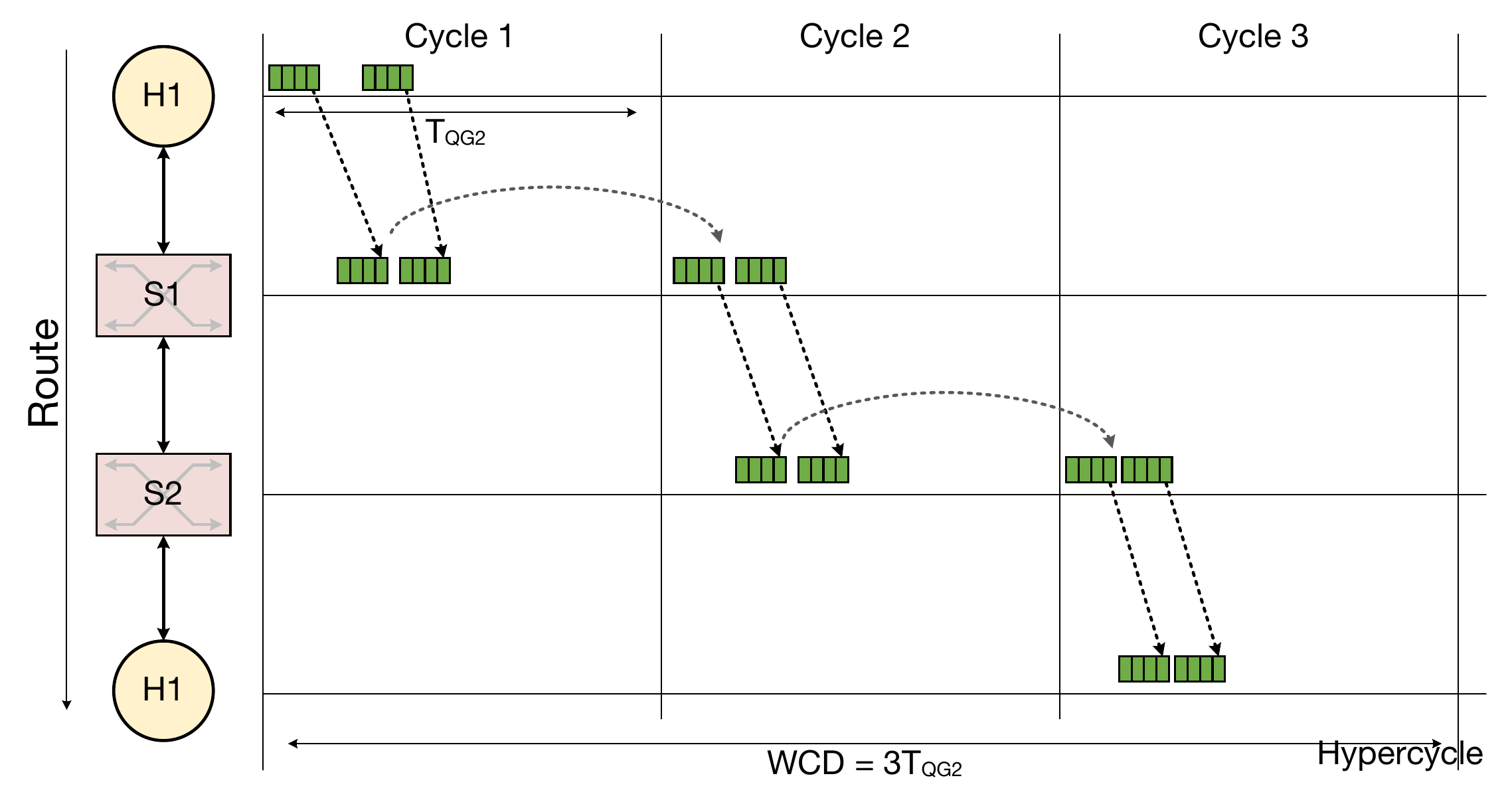}
        \caption{An illustration of flow transmission in a holistic scenario for \texttt{QG}2. The green boxes represent the flows.}
    \label{fig:qg2_wcd}
    \vspace{-0.3cm}
\end{figure}

\vspace{0.3cm}
\textbf{Explanation}: \textit{The frames of the TT flow assigned to the CSQF instance \texttt{QG} experience one cycle of waiting in the queue (due to the \textbf{tolerating queue}) in the worst case (refer to \texttt{QG}1 in Fig.~\ref{fig:mcqf_all_qg_wcd})}. When frames are stored in the \textbf{tolerating queue}, they wait for one cycle before their gate opens for transmission. In contrast, the flows assigned to the CQF instance \texttt{QG} are dequeued in the next cycle, resulting in zero waiting time in the queue (i.e., no queuing delay is imposed on the frame (refer to \texttt{QG}2 and \texttt{QG}3 in Fig.~\ref{fig:mcqf_all_qg_wcd}).

\vspace{0.3cm}
\noindent \textbf{Worst-Case Delay (WCD):}
The worst-case delay of the TT flow ($f_i$) in the Multi-CQF network is given as follows: 
\begin{equation}
\label{eq:wcd}
    \mathrm{WCD}^i = \big( f_i.\phi + \mathrm{SW_{num}^i} + 1 \big) \cdot {T_{\texttt{QG}x}} + (\mathrm{SW_{num}^i} \cdot \mathrm{d_{queue}^i}),
\end{equation}

\noindent where $x = 1,2,3$ for TT flows belonging to \texttt{QG}1, \texttt{QG}2, and \texttt{QG}3 respectively, $f_i.\phi$ represents the offset of the flow $f_i$, $\mathrm{SW_{num}^i}$ is the total number of switches along the route of $f_i$, and $\mathrm{WCD}^i$ is the WCD of the flow $f_i$. To further clarify the WCD formulation, we include Fig.~\ref{fig:mcqf_all_qg_wcd}, \ref{fig:qg1_wcd}, and \ref{fig:qg2_wcd}, which provide a step-by-step visualization of delay accumulation along a flow's route. Fig.~\ref{fig:mcqf_all_qg_wcd} illustrates the queuing delay experienced at each switch for flows in \texttt{QG}x, as well as the overall WCD for flows across different \texttt{QG}s. Fig.~\ref{fig:qg1_wcd} further presents a complete timeline of the WCD for flows belonging to \texttt{QG}1, while Fig.~\ref{fig:qg2_wcd} shows the corresponding delay timelines for flows in \texttt{QG}2.

\vspace{0.3cm}
\textbf{Explanation}: 
\begin{enumerate}
    \item \textit{The offset represented as $f_i.\phi$ is a positive integer value denoting the number of cycles the flow is delayed in the source node. Therefore, $f_i.\phi$ is multiplied with the cycle $T_{\texttt{QG}x}$ to calculate the overall waiting time in the source node (refer to the offset in Fig.~\ref{fig:qg1_wcd})}. 
    \item \textit{$(\mathrm{SW_{num}^i} + 1)$} denotes the total number of hops between the source and the destination. We provide the WCD in terms of $\mathrm{SW_{num}^i}$ to avoid confusion in the calculation of the total number of hops in the route. Depending on the routing protocol, the first link between the source and the switch may or may not be taken into consideration in the total hop count. Therefore, for clarity, we use $(\mathrm{SW_{num}^i} + 1)$ instead of the term hop. The frames are transmitted from one node to the next node in one cycle. Hence, the total time required to traverse through the entire route is denoted as $(\mathrm{SW_{num}^i} + 1)$ multiplied with the cycle ($T_{\texttt{QG}x}$).
    \item \textit{The last term $(SW_{num}^i \cdot \mathrm{d_{queue}^i})$, calculates the queuing delay of the flow in every switch along the route. We take the value of $\mathrm{d_{queue}^i}$ from Eq.~\ref{eq:dqueue}.}
\end{enumerate}

\begin{table}[]
  \centering
  \caption{Summary of Key Notation.}
  \label{tab:notations}
  \resizebox{\columnwidth}{!}{
  \begin{tabular}{|c|l|}
    \hline
    \textbf{Notation} & \textbf{Description} \\
    \hline
    \multicolumn{2}{|c|}{General Terminology} \\ 
    \hline
    GCL & Gate Control List \\
    CQF & Cyclic Queuing and Forwarding \\
    CSQF & Cycle Specific Queuing and Forwarding \\
    Multi-CQF & Multi-Cyclic Queuing and Forwarding \\
    QoS & Quality of Service \\
    TT & Time-Triggered \\
    SA & Simulated Annealing \\
    GA & Genetic Algorithm \\
    GASA & GA-Simulated Annealing \\
    DSK & Domain Specific Knowledge \\
    TAS & Time Aware Shaper \\
    BW & Bandwidth \\
    WCD & Worst-Case Delay \\
    TI & Time Injection \\
    \hline
    \multicolumn{2}{|c|}{TSN Topology} \\ 
    \hline
    \texttt{ESs} & End Stations \\ 
    \texttt{SWs} & Switches \\
    $\mathcal{G}$ & Network Topology \\
    $\mathcal{V}$ & Nodes \\
    $\mathcal{E}$ & Set of Links or Edges \\
    $e_j$ & $j$-th edge number \\
    $\mathcal{F}$ & Set of all TT flows \\
    $\mathcal{|F|}$ & Total Number of TT flows \\
    $f_i$ & The $i$-th TT flow \\
    $f_i.id$ & The TT flow number of the flow $f_i$\\
    $f_i.src$ & The source node of the flow $f_i$ \\
    $f_i.dst$ & The destination node of the flow $f_i$ \\
    $f_i.period$ & The periodicity of the flow $f_i$ in $\mu$s\\
    $f_i.deadline$ & The deadline of the flow $f_i$ in $\mu$s\\
    $f_i.size$ & The frame length of the flow $f_i$ in Bytes (B)\\
    $R$ & Set of all routes of all TT flows \\
    $R_{f_i}$ & Set of \texttt{K-shortest paths} for flow $f_i$ \\
    $r_s$ & The $s$-th route of flow $f_i$ \\
    $T$ & CQF and CSQF cycle \\
    \hline
    \multicolumn{2}{|c|}{Multi-CQF} \\ 
    \hline
    $H$ & Hyperperiod \\
    \texttt{QG} & Multi-CQF Queue Group (can be instance of CQF and/or CSQF) \\
    \texttt{QG}x & Multi-CQF Queue Group number $x$ where $x$ = 1,2,3. \\
    $T_{\texttt{QG}x}$ & Multi-CQF cycle for \texttt{QG}x \\
    $T_{\texttt{QG}1}$ & Multi-CQF cycle for \texttt{QG}1 \\
    $T_{\texttt{QG}2}$ & Multi-CQF cycle for \texttt{QG}2\\
    $T_{\texttt{QG}3}$ & Multi-CQF cycle for \texttt{QG}3 \\
    $\mathrm{d_{queue}^i}$ & Worst-Case Queuing delay of TT flow $f_i$ \\
    $\mathrm{WCD^i}$ & Worst-Case Delay of TT flow $f_i$ \\
    $f_i.\phi$ & Offset of the flow $f_i$ (a positive integer number) \\
    $\mathrm{SW_{num}^i}$ & Total number of switch in the route of TT flow $f_i$ \\
    LCM & Least Common Multiple \\
    GCD & Greatest Common Divisor \\
    $\mathrm{Frame_{num}^i}$ & The total number of frames of TT flow $f_i$ \\
    $T_{min}$ & The minimum cycle \\
    $T_{max}$ & The maximum cycle \\
    $\xi$ & The total delays in the network \\
    $\mathrm{d_{proc}}$ & The processing delay \\
    $\mathrm{d_{prop}}$  & The propagation delay \\
    $\mathrm{sync_{error}}$ & The time synchronization error \\
    $\varphi_{\texttt{QG}x}$ & The percentage of the BW available for \texttt{QG}x \\
    ${\texttt{QG}x}.\mathrm{b}$ & The total number of bits allowed to be transmitted during $T_{\texttt{QG}x}$ \\
    $\mathrm{E2E}(f_i)$ & The total time taken by the flows to reach from the $src$ to the $dst$ \\
    $\mathrm{E2E}\:(f_{i, r_{s}})$ & The end-to-end delay of flow $f_i$ with route $r_s$. \\
    $l,m,n$ & Configurable BW percentage allocated to the \texttt{QG}s \\
    $\alpha, \beta$  & Optimization parameters \\
    \hline
  \end{tabular}}
\end{table}

\section{Multi-CQF Problem Formulation}
\label{sec:problem}
In this section, we discuss the constraints and the problem formulation of Multi-CQF. 

\vspace{0.2cm}
\noindent \textbf{Deadline:} To successfully schedule a TT flow, it must meet its deadline. Therefore, for all TT flows, $\forall f_i \in $ $\mathcal{F}$, 
\begin{equation}
\label{eq:deadline_Constraint_one}
    C_1: \; \mathrm{WCD}^i \leq f_i.deadline,
\end{equation}
\noindent where the $\mathrm{WCD}^i$ is given in Eq.~\ref{eq:wcd}.

\subsection{Core Constraints}
\label{subsec:constraints}
The proper functioning of cyclic shapers requires a sufficiently large cycle to ensure that all frames received by the node in one cycle are fully transmitted during the next cycle~\cite{boyer_2024_cqf_cycle}. Therefore, it is crucial to define the proper constraints for the Multi-CQF cycle. It is important to note that determining the optimal cycle is itself an optimization problem, as discussed in \cite{boyer_2024_cqf_cycle}. In the context of Multi-CQF, this challenge becomes more complex, as we must determine three distinct cycles, one for each \texttt{QG}x (\texttt{QG}1, \texttt{QG}2, and \texttt{QG}3).

\vspace{0.5em}
\noindent\textbf{Hyperperiod:} All ports of the TSN switch are configured to operate cyclically. This cycle repeats itself and is known as the Hyperperiod $(H)$ or the scheduling cycle. In short, the configuration of the network repeats over the $H$, and is defined as:
\begin{equation}
    C_2: \; H = \mathrm{LCM}(f_{i}.period), \; \forall f_{i} \in F, 
\end{equation}
\noindent where $f_{i}.period$ denotes the periodicity of the TT flow $f_i$. 

\vspace{0.5em}
\noindent\textbf{Frame Number:} The total number of TT flow frames transmitted within the $H$ is given as:
\begin{equation}
\label{eq:frame_number}
    C_3: \; \mathrm{Frame_{num}^i} = \frac{H}{f_i.period}.
\end{equation}
\noindent Eq.~\ref{eq:frame_number} gives the total number of cycles occupied by the frames of flow $f_i$ in the $H$.

\vspace{0.5em}
\noindent\textbf{Minimum Cycle ($T_{min}$)}: The cycle $T_{\texttt{QG}x}$ should be long enough to transmit the largest frame size in the network. Therefore, the minimum cycle $T_{min}$ is defined as: 
\begin{equation}
    C_4: \; T_{min} \geq \frac{\mathrm{max}(f_{i}.size) \cdot 8}{\mathrm{BW}} + \xi, \; \forall f_{i} \in F, 
\end{equation}
where $\xi$ is a constant value for the delays which compensates for the synchronization error, and other delays in the network.
\begin{equation}
\xi = \mathrm{d_{proc}} + \mathrm{d_{prop}} + \mathrm{sync_{error}},
\end{equation}
\noindent where $\mathrm{d_{proc}}$ is the processing delay, $\mathrm{d_{prop}}$ is the propagation delay, and $\mathrm{sync_{error}}$ is the time synchronization error.

\vspace{0.5em}
\noindent\textbf{Maximum Cycle ($T_{max}$)}: The maximum cycle denoted as $T_{max}$ is the largest possible cycle, where all the flows periodicities should be divisible by the cycle. Therefore, it is calculated as the Greatest Common Divisor (GCD) of all the TT flow periods\cite{burst_aware_mcqf},\cite{mss_transactions}, \cite{frer_cqf_toades} and given as:
\begin{equation}
    C_5: \; T_{max} \leq \mathrm{GCD}(f_{i}.period), \; \forall f_{i} \in F. 
\end{equation}

\noindent \textbf{Possible Cycle:} Now, we discuss the constraints for the Multi-CQF cycle. Based on the proposed constraints, we identify all the possible cycles and then select three different cycles from these available options. 

\noindent Below constraint ensures that $H$ is divisible by all $T_{\texttt{QG}x}$ such that we have a whole number of cycles in $H$.
\begin{equation}
\label{eq:c6}
    C_6: \; H \; \% \; T_{\texttt{QG}x} = 0,
\end{equation}
where \% represents the \texttt{mod} operator.

\vspace{0.5em}
\noindent We apply the below constraint to ensure that all frames of the flow are generated during the start of the cycle within the $H$.
\begin{equation}
\label{eq:c7}
    C_7: \; f_{i}.period \; \% \; T_{\texttt{QG}x} = 0,  \; \forall f_{i} \in F. 
\end{equation}

\vspace{0.5em}
\noindent\textbf{\texttt{QG} Cycle:} We select three different $T_{\texttt{QG}x}$ between $T_{min}$ and $T_{max}$.
\begin{equation}
\label{eq:qg1}
    C_8: \; T_{min} \leq T_{\texttt{QG}x} \leq T_{max}.
\end{equation}
Equation \ref{eq:qg1} ensures that the selected cycle is larger or at least equal to the minimum cycle. Furthermore, the selected $T_{\texttt{QG}x}$ should be less than the maximum cycle.
\begin{equation}
\label{eq:qg2}
    C_9: \; T_{\texttt{QG}1} < T_{\texttt{QG}2} < T_{\texttt{QG}3}.
\end{equation}
Equation \ref{eq:qg2} ensures that the cycle for \texttt{QG}1 is smaller than \texttt{QG}2 and \texttt{QG}3. Similarly, the cycle for \texttt{QG}2 is smaller than \texttt{QG}3.
\begin{equation}
\label{eq:nu1}
    C_{10}: \; T_{\texttt{QG}2} \; \% \; T_{\texttt{QG}1} = 0  \; \; \land \;\; T_{\texttt{QG}3} \; \% \; T_{\texttt{QG}2} = 0.
\end{equation}
$C_{10}$ ensures that the $T_{\texttt{QG}x}$ of the \texttt{QG}s are an integral multiple of each other such that the opening and closing of the gates for different \texttt{QG}s are aligned. Equation \ref{eq:nu1} checks that $T_{\texttt{QG}2}$ is fully divisible by $T_{\texttt{QG}1}$ and $T_{\texttt{QG}3}$ is fully divisible by $T_{\texttt{QG}2}$ and $T_{\texttt{QG}1}$ to satisfy the integral multiple requirement. It is important to select the different \texttt{QG} cycles as an integral multiple of each other to prevent over-subscription of the $\mathrm{BW}$. For more details, we direct the reader to refer to \cite{norman}. 

\vspace{0.5em}
\noindent\textbf{BW Distribution:} The flows sent by all \texttt{QG} on each link/edge $e_j$ must be less than the allocated BW to the \texttt{QG}. 
Additionally, all \texttt{QG} together must transmit flows less than the total BW of the link.
\begin{equation}
\label{eq:bandwidth_constraint}
C_{11}: \; \sum_{f_{i} \in |\texttt{QG}x|} (f_i.size) \cdot \frac{e_j}{f_i.period} \leq \varphi_{\texttt{QG}x} \cdot \mathrm{BW},
\end{equation}
\begin{equation}
  e_j =
    \begin{cases}
      1  & , \mathrm{j^{th} \: edge} \in  R_{f_i},\\
      0 & , \mathrm{otherwise},
    \end{cases}       
\end{equation}
\noindent where $\varphi_{\texttt{QG}x}$ is the percentage of the BW available for \texttt{QG}x. In this paper, we consider the $\varphi_{\texttt{QG}x}$ as configurable by the system engineer. Depending on the application requirements, the BW distribution among \texttt{QG}s can be adjusted accordingly. In this work, we treat the BW distribution as a predefined input parameter. Following the approach in \cite{mcqf_paul}, we assign a larger share of BW to \texttt{QG}1, followed by \texttt{QG}2, and then \texttt{QG}3. This prioritization is based on the observation that more flows are typically mapped to \texttt{QG}1, followed by \texttt{QG}2 and \texttt{QG}3, as discussed in detail in Section~\ref{sub:flow_to_qg_mapping}. 

\vspace{0.5em}
\noindent \textbf{Bits transmitted per Cycle:}
Given the BW distribution, we compute the total number of bits that can be transmitted during each cycle $T_{\texttt{QG}x}$ in our algorithm. This ensures that no more bits are transmitted than what the cycle can accommodate. The total number of bits permitted for transmission in each cycle is defined by the following equation.
\begin{equation}
\label{eq:group_cycle_bits}
{\texttt{QG}x}.\mathrm{b} = (\varphi_{\texttt{QG}x} \cdot \mathrm{BW}) \cdot T_{\texttt{QG}x},
\end{equation}
\noindent where, ${\texttt{QG}x}.\mathrm{b}$ denotes the total number of bits that can be transmitted during a single cycle $T_{\texttt{QG}x}$. The unit of ${\texttt{QG}x}.\mathrm{b}$ is in bits. 

\vspace{0.5em}
\noindent\textbf{Time Injection (TI):} TI is the time offset that indicates when TT flows are transmitted from their source node. This constraint ensures that each TT flow is sent within its periodicity to prevent the accumulation of frames at the source node: 
\begin{equation}
\label{eq:ti}
C_{12}: \; \forall f_{i} \in F, (\; f_i.\phi \geq 0) \cap \Big(f_i.\phi < \left\lfloor\frac{f_i.period}{T_{\texttt{QG}x}}\right\rfloor\Big).
\end{equation}

\noindent\textbf{Optimization Function:} The optimization goal of this paper is to minimize the end-to-end delay ($\mathrm{E2E}$) of the flows. $\mathrm{E2E}$ is defined as the total time taken by the flows to reach from the source to the destination. It is important to note that $\forall i \in $ $\mathcal{F}$, $\mathrm{E2E}(f_i)$ $\leq$ WCD($f_i$).
\begin{align} 
\label{eq:e2e}
\min_{{f}_{i}\in \mathcal{F}}& \frac{1}{|\mathcal{F}|}\sum_{r_s\in R_{f_i}}\mathrm{E2E}\:(f_{i, r_{s}})\,, \\
s.t. & \: C_{1}  \: \mathrm{to} \: C_{12} \: \mathrm{is \: satisfied,} \notag 
\end{align}
\noindent where, $\mathrm{E2E}\:(f_{i, r_{s}})$ represents the end-to-end delay of flow $f_i$ when taken the route $r_s$.

\begin{figure*}[t!]
    \centering    
    \includegraphics[scale=0.42, trim={0.3cm 0.6cm 0.2cm 0.5cm}, clip]{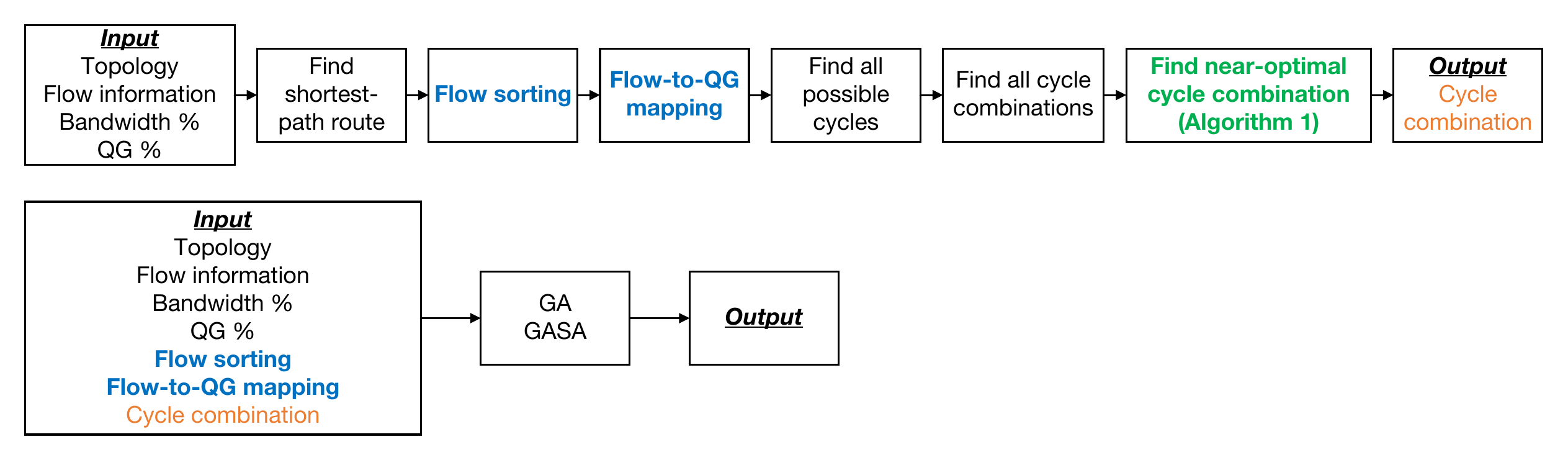}
    \caption{Overview of the proposed Multi-CQF configuration workflow. The diagram illustrates the steps, including input parameters, pre-processing (flow sorting and flow-to-\texttt{QG} mapping), cycle calculation, and the input to GA and GASA algorithms.}
    \vspace{-0.2cm}
\label{fig:flowchart}
\end{figure*}

\section{Implementation}
\label{sec:implementation}
Fig.~\ref{fig:flowchart} presents a block diagram outlining the overall workflow of our approach. This figure illustrates the key components of our model, highlighting the input parameters, the pre-processing steps including flow sorting and \texttt{QG} mapping, the cycle calculation, and the inputs passed to the GA and GASA algorithms.

In this section, we first describe the flow-to-\texttt{QG} mapping choice and then introduce optimizations using DSK to balance configuration quality with computational efficiency. The Multi-CQF configuration problem is known to be NP-hard~\cite{mcqf_paul}. To reduce this complexity, we employ flow sorting and assign flows to specific \texttt{QG}x before Multi-CQF configuration. We begin by outlining various flow-to-\texttt{QG} mapping mechanisms and discuss how incorporating DSK reduces the search space and improves convergence efficiency.

\vspace{-0.5cm}
\subsection{Flow sorting and flow-to-\texttt{QG} mapping}
\label{sub:flow_to_qg_mapping}
Flow sorting and flow-to-\texttt{QG} mapping significantly influence the overall schedulability of the network, as shown in Section~\ref{sec:results}. Unlike CQF and CSQF, Multi-CQF involves both sorting flows and distributing them across different \texttt{QG}s before scheduling. In our framework, flow sorting and mapping serve two primary purposes:
(1) assigning flows to appropriate \texttt{QG}s based on their characteristics, and (2) determining the scheduling order within each \texttt{QG}. In this paper, we allocate $l\%$ of the flows with the shortest deadlines to \texttt{QG}1, the next $m\%$ to \texttt{QG}2, and the remaining $n\%$ to \texttt{QG}3, where $l,m,$ and $n$ are configurable. The specific values used for $l$, $m$, and $n$ are discussed below. 

This approach is motivated by the fact that flows with tighter deadlines benefit from smaller cycles, and thus should be assigned to \texttt{QG}s with shorter cycles. Since the WCD depends on the cycle, pre-assigning flows to \texttt{QG}s based on deadline helps reduce the configuration search space. Our algorithm does not need to test all flows across all three \texttt{QG}s, thus reducing the search space for the configuration leading to a faster convergence. This flow mapping is performed before the scheduling begins, and the grouped flows are then passed to the GA and GASA algorithms. To study the impact of different sorting and mapping strategies in detail, we consider the following approaches:
\begin{enumerate}
    \item \textbf{Deadline-Based Mapping (DBM):} In Deadline-Based Mapping (DBM), we sort the flows by deadline in ascending order. The first 50\% of the flows are assigned to \texttt{QG}1, the next 30\% to \texttt{QG}2, and the remaining 20\% to \texttt{QG}3. Since \texttt{QG}1 has the smallest cycle, it is best suited for flows with the tightest deadlines to reduce WCD. If we assign such flows to \texttt{QG}2 or \texttt{QG}3, which have larger cycles, it increases the possibility of deadline violations.
    \item \textbf{Periodicity-Based Mapping (PBM):} Here we sort the flows by periodicity in ascending order. The most frequent (smallest periodicity) flows comprising the first 50\% are assigned to \texttt{QG}1, followed by 30\% to \texttt{QG}2 and the remaining 20\% to \texttt{QG}3. Flows with smaller periodicity benefit from assignment to smaller cycle \texttt{QG}, as they need to be scheduled more often within the $H$.
    \item \textbf{Random Mapping (RM):} In the random mapping strategy, flows are assigned to \texttt{QG}1, \texttt{QG}2, and \texttt{QG}3 arbitrarily, without considering any specific flow characteristics. We assign 50\% flows to \texttt{QG}1, 30\% flows to \texttt{QG}2, and the remaining 20\% flows to \texttt{QG}3.
\end{enumerate}

\begin{figure}[t!]
    \centering    
    \includegraphics[scale=0.5, trim={0.8cm 0.6cm 0.9cm 0.5cm}, clip]{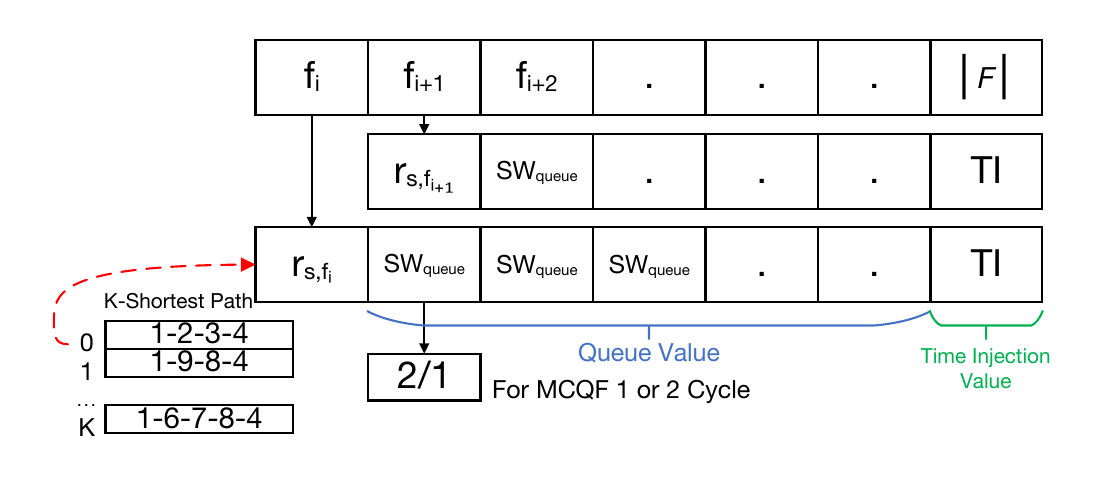}
    \caption{State Vector representation of  the GA and GASA model.}
    \vspace{-0.2cm}
\label{fig:chromosome}
\end{figure}

\vspace{-0.2cm}
\subsection{Search Strategy}
\label{sub:ss}
The search space for the CQF network is $\Pi_{i=1}^{\mathcal{|F|}}\frac{f_i.period}{T}$. When the $T$ is small in CQF network the search space becomes large requiring high computational time. By using flow sorting and flow-to-\texttt{QG}x assignment, we optimize the search space for the Multi-CQF network to $\Pi_{i=1}^{|\mathcal{F}_{\texttt{QG}1}|}\frac{f_i.period}{\mathcal{F}_{\texttt{QG}1}}$ + $\Pi_{i=1}^{|\mathcal{F}_{\texttt{QG}2}|}\frac{f_i.period}{T_{\texttt{QG}2}}$ + $\Pi_{i=1}^{|\mathcal{F}_{\texttt{QG3}}|}\frac{f_i.period}{T_{\texttt{QG}3}}$, where $|F_{\texttt{QG1}}|$, $|\mathcal{F}_{\texttt{QG2}}|$, and $|\mathcal{F}_{\texttt{QG3}}|$ denote the total number of flows in \texttt{QG}1, \texttt{QG}2, and \texttt{QG}3 respectively. Furthermore, $\mathcal{F}_{\texttt{QG}1} \cap \mathcal{F}_{\texttt{QG}2} = \emptyset, \quad \mathcal{F}_{\texttt{QG}2} \cap \mathcal{F}_{\texttt{QG}3} = \emptyset, \quad \mathcal{F}_{\texttt{QG}1} \cap \mathcal{F}_{\texttt{QG}3} = \emptyset$, $\quad \mathcal{F}_{\texttt{QG}1} \cup \mathcal{F}_{\texttt{QG}2} \cup \mathcal{F}_{\texttt{QG}3} \subseteq \mathcal{F}.$

\vspace{-0.3cm}
\subsection{Cycle Combinations}
\label{sec:cycle_model}
We determine an efficient cycle combination using a heuristic search mechanism. This model operates independently and is not integrated into the SA, GA, or GASA algorithms. As shown in Fig.~\ref{fig:flowchart}, the cycle combination model is a separate module whose output is provided as input to the GA and GASA algorithms. Algorithm~\ref{algo:alg1} outlines the procedure to compute the efficient cycle combination. In line 1, we compute the shortest-path routing for the TT flows, followed by flow sorting and flow-to-\texttt{QG} mapping. In line 2, we apply the constraints to determine feasible values of $T_{\texttt{QG}x}$ and store them in a list. Line~3 constructs all possible cycle combinations $(T_{\texttt{QG}1}, T_{\texttt{QG}2}, T_{\texttt{QG}3})$ across the different \texttt{QG}s and stores them in array $K$. We initialize the current best schedulability score $S$ to 0 (line 4) and set the best cycle combination $T_{\text{best}}$ as an empty set (line 5). Lines 6 to 13 iterate over all candidate cycle combinations in $K$, checking each against the deadline and BW constraints and computing the resulting schedulability score $S'$. If the current cycle combination yields a higher schedulability score, we update $S$ and store the corresponding cycle combination as the new $T_{\text{best}}$ (lines 9-11). Finally, line~14 returns the best performing cycle combination $T_{\text{best}}$. \textit{It is important to note that our model may not provide the optimal cycle combination as this can only be guaranteed using a well-defined optimization function. Our goal is to find an efficient cycle combination for a given TC, instead of using random cycle combination.}

\begin{algorithm}[!t]
  \caption{Cycle Combination}
  \label{algo:alg1}
  \begin{algorithmic}[1]
    \Require $\mathcal{G}(\mathcal{V}, \mathcal{E})$, $\mathcal{F}$, \text{BW}, $\varphi_{\texttt{QG}x}$
    \Ensure Best Cycle combination, $T_{best}$ $\leftarrow$ $(T_{\texttt{QG}1},T_{\texttt{QG}2}, T_{\texttt{QG}3})$ 
    \State Find Shortest-Path routing, perform flow sorting and flow-to-\texttt{QG} mapping (e.g., DBM, PBM, RM).
    \State Use the constraints to find possible $(T_{\texttt{QG}x})$ values and store in a list. 
    \State Create all possible combinations of $(T_{\texttt{QG}x})$ values $(T_{\texttt{QG}1},T_{\texttt{QG}2}, T_{\texttt{QG}3})$ for Multi-CQF using the discussed constraints and store in array $K$.
    \State $S \gets 0$ \Comment{Current best schedulability score}
    \State $T_{\text{best}} \gets \emptyset$ 
    \For{$k \in K$} \Comment{Loop through all possible combinations}
      \State Pass $k$ through deadline and \text{BW} constraints
      \State $S' = \frac{\text{Total Number of Scheduled Flows}}{|\mathcal{F}|}$ \Comment{Compute schedulability score $S'$}
      \If{$S' > S$}
        \State $S \gets S'$
        \State $T_{\text{best}} \gets k$
      \EndIf
    \EndFor
    \State \Return $T_{\text{best}}$
  \end{algorithmic}
\end{algorithm}

\subsection{Genetic Algorithm}
\label{sec:ga}
\noindent \textbf{Fitness Function:} The fitness function evaluates the quality of a solution for the given problem. In this paper, the fitness function of the proposed algorithm aims to minimize the equation given below. 

\begin{align} 
  \label{eq:fitness}
  \mathrm{min} \; \frac{1}{\mathcal{|F|}}\sum_{f_i\in \mathcal{F}}\frac{\mathrm{E2E}\:(f_{i, r_{s}})}{f_{i}.deadline} + \alpha\cdot\mathrm{C1} + \beta\cdot\mathrm{C11}.
\end{align}
\noindent where $f_{i, r_{s}}$ is the route $r_s$ taken by $f_i$, \texttt{C1} and \texttt{C11} denotes the constraint one (\ref{eq:deadline_Constraint_one}) and constraint eleven (\ref{eq:bandwidth_constraint}) violations respectively. The penalty offered for \texttt{C1} is $\alpha$, and the penalty for violating \texttt{C11} is $\beta$. After normalizing, the fitness value falls between 0 and 1 when all constraints are satisfied and all the flows are scheduled. To balance the trade-off between \texttt{C1} and \texttt{C11} violations, we systematically analyze the impact of different combinations of $\alpha$ and $\beta$ values under varying network sizes and flow densities. The detailed results of this evaluation are presented in Section~\ref{sec:results}, where we include a heatmap to illustrate the influence of $\alpha$ and $\beta$ on scheduling feasibility. Based on this study, we set $\alpha = 2$ and $\beta = 2$ for all subsequent experiments, as this combination offers a balanced performance across the tested scenarios (refer to Fig.~\ref{fig:alpha_beta_mcqf} in Section~\ref{sec:results}).

\vspace{0.5em}
\noindent \textbf{Routing:} We find the K-shortest paths for each flow with the given source and destination node. For the routing we used the NetworkX\footnote{NetworkX:\url{https://networkx.org/documentation/stable/tutorial.html}} library to find the K-shortest path. The algorithm first examines the shortest path, and then checks the next potential route from the set of available routes if the currently selected route cannot schedule the flow. 

\vspace{0.5em}
\noindent \textbf{Time Injection:}
With the introduction of TI, the WCD of a TT flow is given in (Eq.~\ref{eq:wcd}). In our model, TI is a positive integer value ranging from 0 to $\left\lfloor\frac{f_i.period}{T_{\texttt{QG}x}}\right\rfloor - 1$. 

\begin{algorithm}[!t]
\caption{Genetic Algorithm} 
\label{alg:ga}
\begin{algorithmic}[1]
    \State Initialize population $P$ \Comment{initial random Multi-CQF solution $P$}
    \State Initialize fitness $F \gets \Call{E2E\_Objective}{P}$
    \State Initialize elites $E \gets P$ \Comment{$E$ is the entire initial solution}
    \While{\textbf{not} termination\_condition }
        \State $all\_pairs \gets \Call{select\_parents}{E}$ \Comment{select solution pairs}
        \For{\textbf{each} ($p1,p2$) \textbf{in} $all\_pairs$} \Comment{$p1$ and $p2$ are two solutions}
            \State $child1, child2 \gets \Call{Crossover}{p1,p2}$
            \State $child1 \gets \Call{Mutate}{child1}$ \Comment{randomly change one value of the $child$ sub-vector}
            \State $child2 \gets \Call{Mutate}{child2}$
            \State Population $P$ add $child1,child2$  \Comment{$child$ is now part of the whole solution $P$}
         \EndFor
        \State $Q \gets \Call{E2E\_Objective}{P}$ \Comment{objective function value}
        \State $P \gets \Call{Sort}{P, \text{key}=Q}$ \Comment{sort the solution by fitness score}
        \State $E \gets P[:\text{len}(P)//2]$ \Comment{remove half of the weakest solutions}
        \State termination\_condition $\gets \Call{IsTerminate}{E}$
        \EndWhile
    \State \Return $E$ 
\end{algorithmic}
\end{algorithm}

\begin{algorithm}[!t]
\caption{Crossover and Mutation Function}
\label{alg:pseudo_crossover}
\begin{algorithmic}[1]
\Function{Crossover}{$parent1, parent2$} \Comment{$parent$ is one of the Multi-CQF solutions}
\State Initialize $child1,child2$ 
    \For{\textbf{each} ($gene1, gene2$) \textbf{in} ($parent1, parent2$)}
    \State $child1$ add \Call{ChooseRandom}{$gene1,gene2$}
    \State $child2$ add \Call{ChooseRandom}{$gene1,gene2$}
    \EndFor
    \State \Return $child1,child2$
\EndFunction
\Function{Mutation}{$individual$}
    \State $\Call{Simulated Annealing}{}$ \Comment {Run SA search method for GASA}
    \State $prob \gets \Call{Random}{0,1}$
    \If {$prob < MutationRate$}
        \State \Call{Replace Gene}{$individual$}
        \State \Call{Simulated Annealing}{$individual$} \Comment{only replace the gene if it is better, and replace multiple genes} 
    \EndIf
    \State \Return $individual$
\EndFunction
\end{algorithmic}
\end{algorithm}

\vspace{0.5em}
\noindent \textbf{Chromosome Representation:}
\label{subsec:state_vector}
GA is a population-based method using a set of individuals, each represented by numerical values known as chromosomes. In our GA model, chromosomes are represented by a state vector as shown in Fig.~\ref{fig:chromosome}. The vector consists of $|\mathcal{F}|$ blocks, where each block represents a TT flow denoted as $f_i$, $f_{i+1}$...$f_{|\mathcal{F}|}$. Each flow block $f_i$ includes a sub-vector detailing the index number of possible routes, $R_{f_i}$. The route $R_{f_{i}}$ contains all the possible routes for the flow $f_i$. The sub-vector specifies: (1) the index of the K\textsuperscript{th} possible path for the flow, (2) the number of \texttt{SW}s in the route, and (3) the last entry contains the TI values (a positive integer) indicating the number of cycle the flow is delayed in the $src$ node. Algorithm~\ref{alg:ga} outlines the parent selection process. In lines 1-3, we initialize the population, define the fitness function to minimize end-to-end delay, and set the elite group as the initial population. New solutions are created by pairing individuals and selecting parents (line 5). Elite selection (line 13) chooses the top $N$ most fit individuals based on their \texttt{fitness score} (refer to~\ref{eq:fitness}). Line 7 performs \texttt{crossover} between parents \texttt{p1} and \texttt{p2} to generate a new Multi-CQF solution as a state vector with queue mapping. Lines 8-9 mutate \texttt{child1} and \texttt{child2} to diversify the population, altering one or more values in the state vector. Line 14 updates the \texttt{elites} by retaining the best individuals and removing the weakest. Line 15 sets the termination condition, which could be the number of iterations or a time limit. For crossover and mutation given in Algorithm~\ref{alg:pseudo_crossover}, we use two pairing mechanisms: (1) randomly selecting the complete flow configuration from each parent and (2) a randomized method that mixes the internal configuration of individual flows by selecting parts from each parent. In line 4 of Algorithm~\ref{alg:pseudo_crossover}, \texttt{gene1} and \texttt{gene2} represent the single value of the state sub-vector. Line 10 gives the \texttt{mutation} function where the mutation frequency is based on the \texttt{mutation rate}. In the GASA algorithm, we set the mutation based on the SA mechanism as given in line 15.

\begin{figure*}[!t]
    \centering
    \begin{minipage}[b]{0.22\textwidth}
        \centering
        \includegraphics[scale=0.2, trim={0.4cm 0.4cm 0.4cm 0.4cm}, clip]{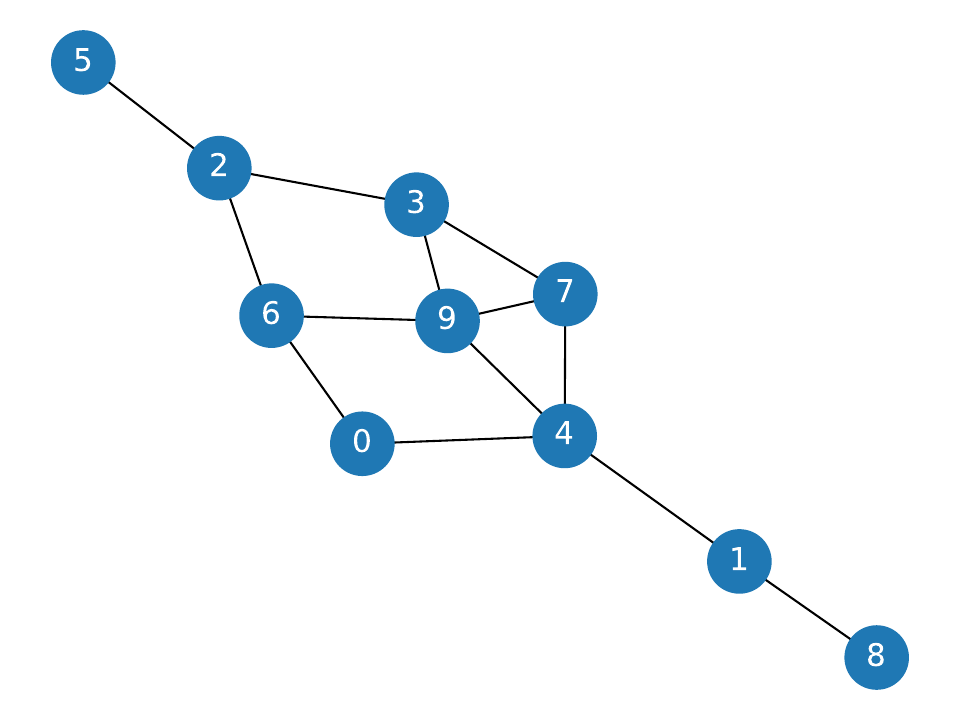}
        \subcaption{}
    \label{fig:erg}
    \end{minipage}
    \begin{minipage}[b]{0.22\textwidth}
        \centering
        \includegraphics[scale=0.2, trim={1cm 0.6cm 1cm 0.6cm}, clip]{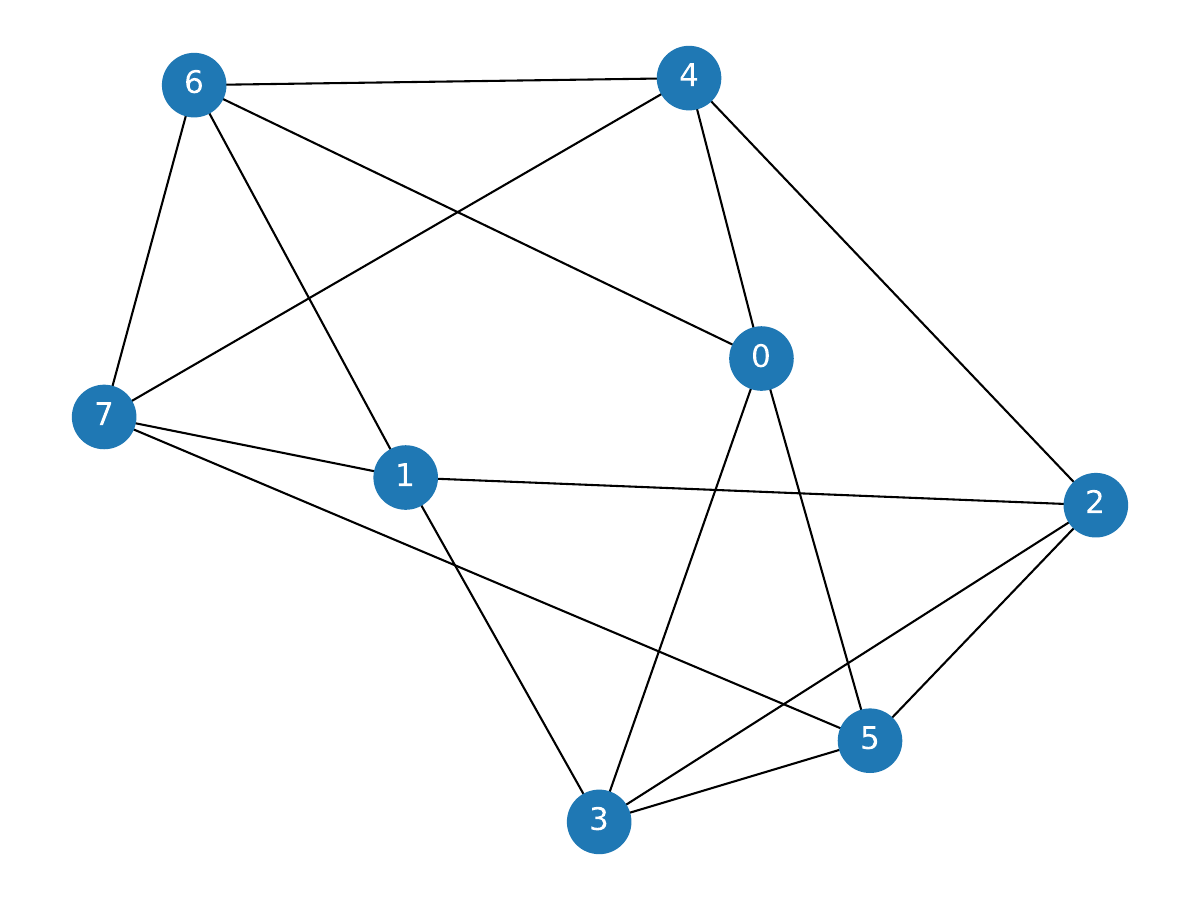}
        \subcaption{}
    \label{fig:rrg}
    \end{minipage}
    \begin{minipage}[b]{0.22\textwidth}
        \centering
        \includegraphics[scale=0.2, trim={1cm 0.4cm 1cm 0.6cm}, clip]{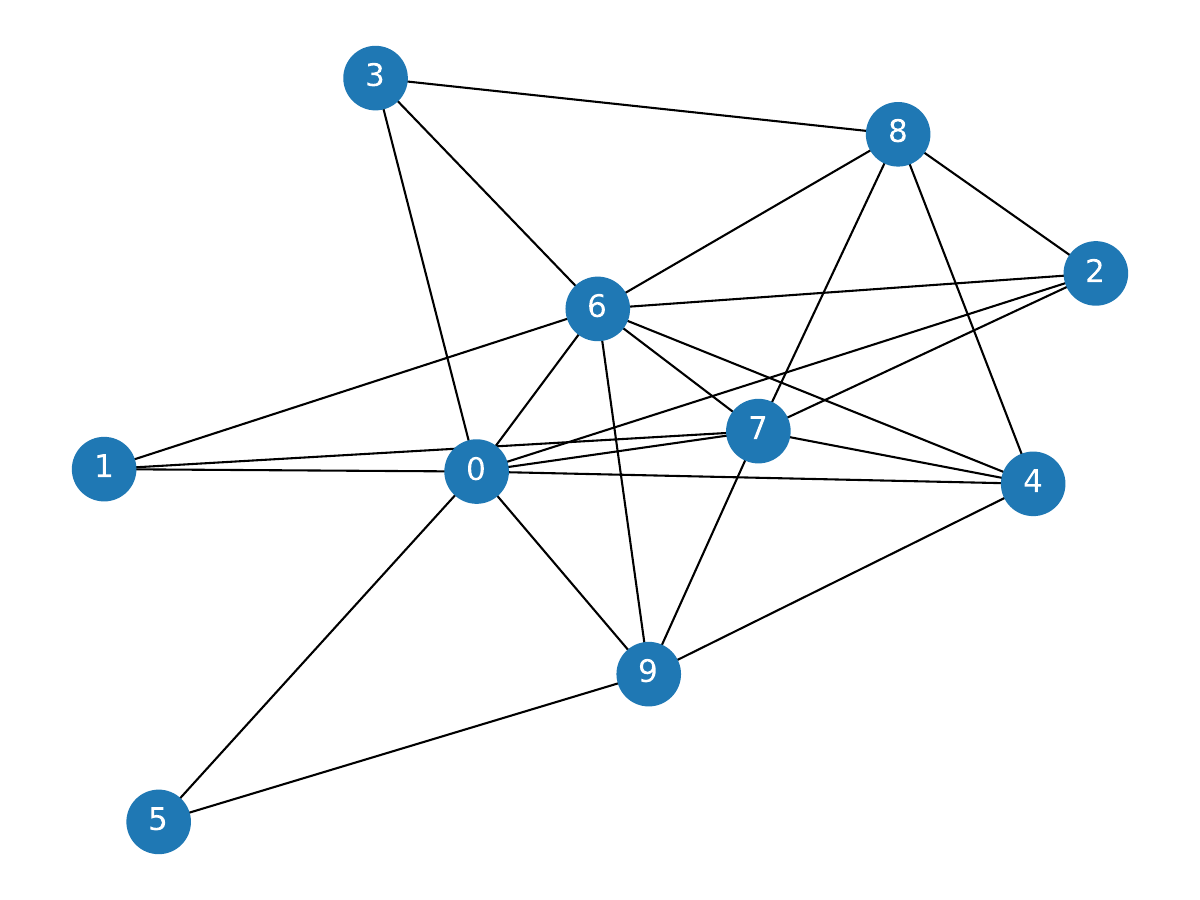}
       \subcaption{}
    \label{fig:bag}
    \end{minipage}
    \begin{minipage}[b]{0.22\textwidth}
        \centering
    \includegraphics[scale=0.11, trim={19cm 2.6cm 19cm 0.6cm}, clip]{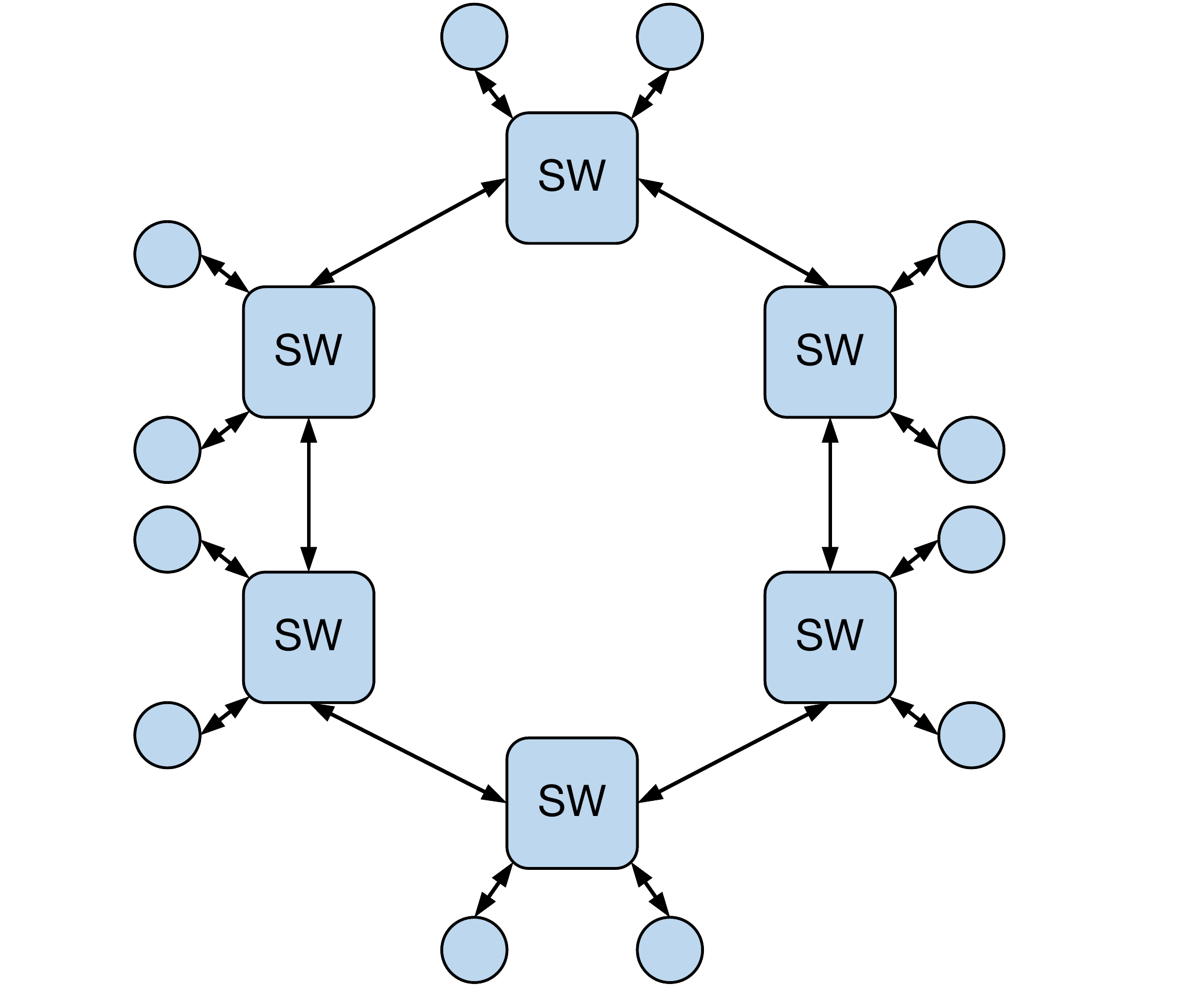}
    \subcaption{}
    \label{fig:ring}
    \end{minipage}
    \caption{Topologies under test: (a) Erdos-Renyi Graph [ERG (10,0.3)]~\cite{deepscheduler}, (b) Random Regular Graph [RRG (8,4)]~\cite{deepscheduler}, (c) Barabasi-Albert Graph [BAG (10,4)]~\cite{deepscheduler}, and (d) Ring (for Ring topology, we tested our model with varying network sizes, ranging from 26 to 156 nodes).}
    \label{fig:erg_rrg_bag_ring}
\end{figure*}

\begin{figure*}[!t]
    \centering
    \begin{minipage}[b]{0.48\textwidth}
    \centering
    \includegraphics[scale=0.42, trim={0.4cm 0.4cm 0.5cm 0.4cm}, clip]{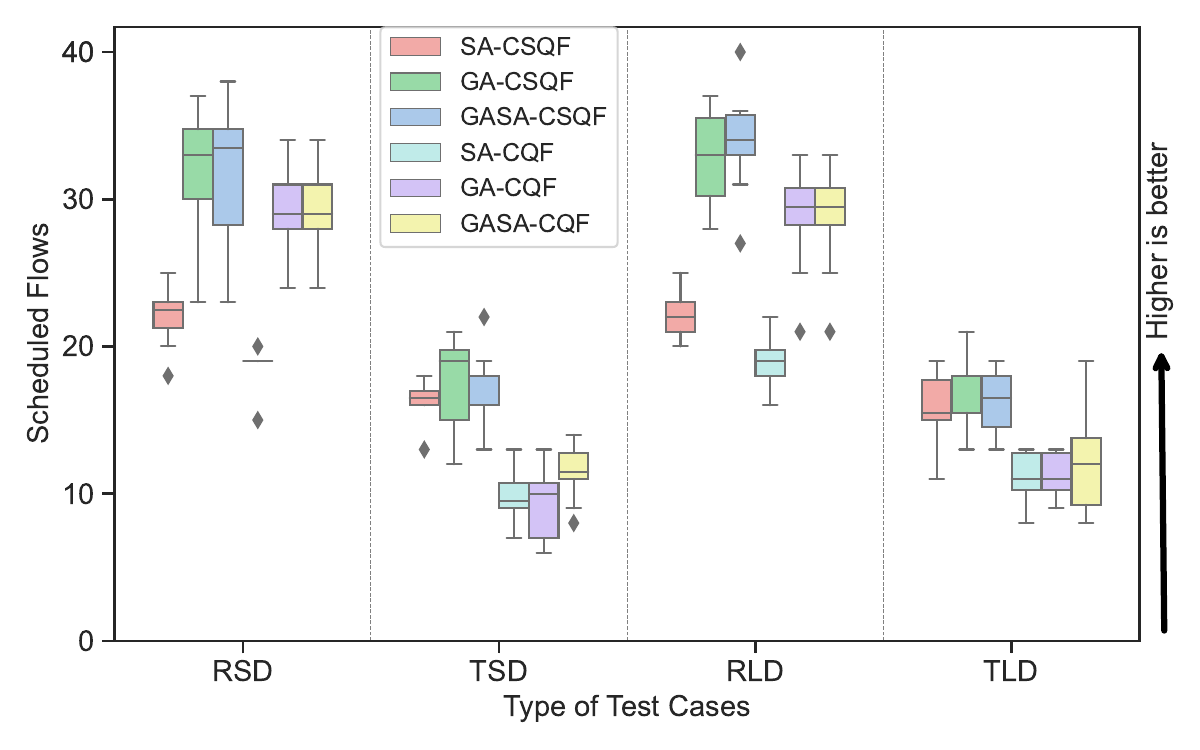}
    \subcaption{Schedulability}
    \label{fig:cqf_csqf_ER_boxplot_schedule}
    \end{minipage}
    \begin{minipage}[b]{0.48\textwidth}
    \centering
    \includegraphics[scale=0.42, trim={0.4cm 0.4cm 0.5cm 0.4cm}, clip]{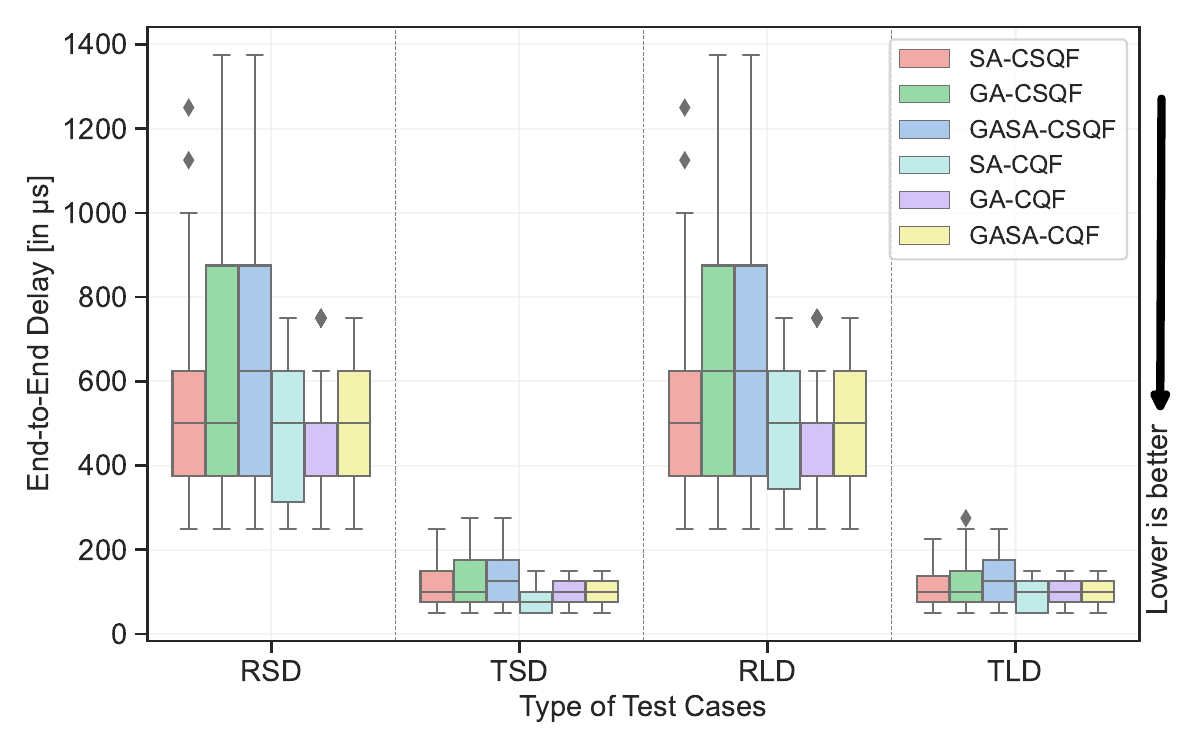}
    \subcaption{WCD} 
\label{fig:cqf_csqf_ER_boxplot_delay}
    \end{minipage}
    \caption{Comparison of CQF and CSQF for ERG Topology (Fig.~\ref{fig:erg}) with 100 Mbps BW. For the Tight test case, $T = 25\mu s$ for both CQF and CSQF, while for the Relaxed test case, $T = 125\mu s$.} 
    \label{fig:cqf_csqf_ER}
\end{figure*}

\begin{figure*}[!t]
    \centering
    \begin{minipage}[b]{0.48\textwidth}
    \centering
    \includegraphics[scale=0.42, trim={0.4cm 0.4cm 0.5cm 0.4cm}, clip]{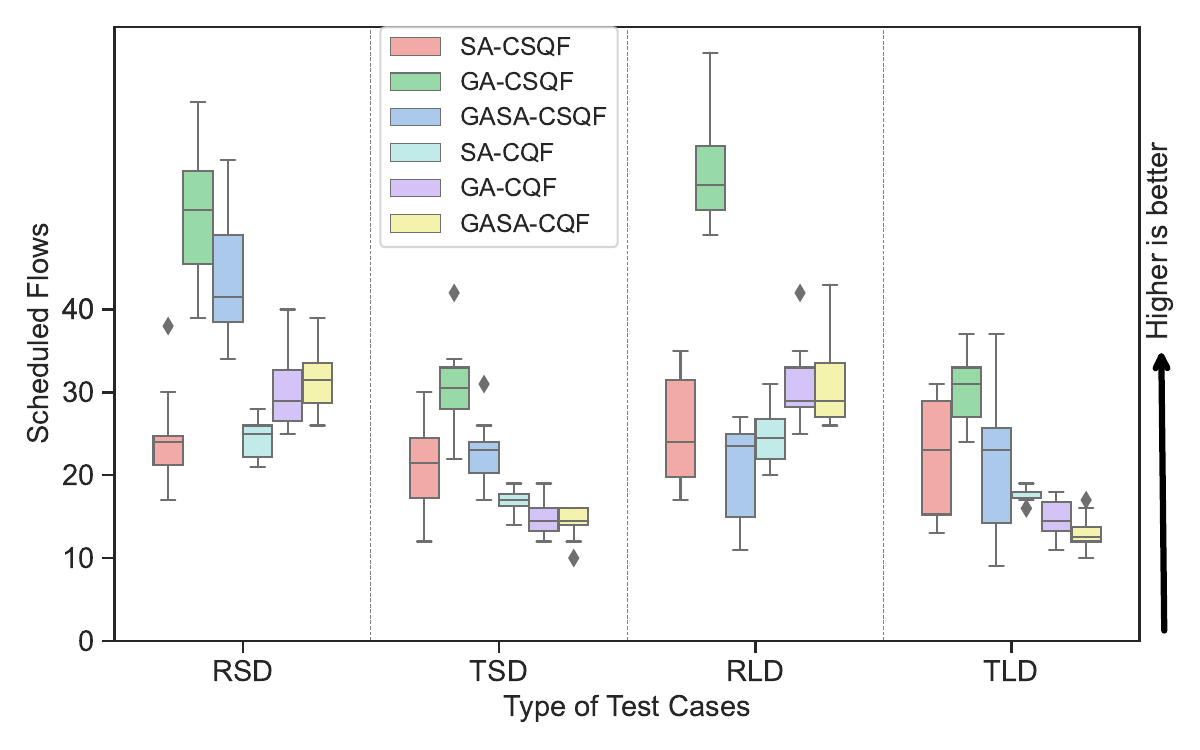}
    \subcaption{Schedulability}
    \label{fig:box_plot_cqf_csqf_RRG_schedule}
    \end{minipage}
    \begin{minipage}[b]{0.48\textwidth}
    \centering
    \includegraphics[scale=0.42, trim={0.4cm 0.4cm 0.5cm 0.4cm}, clip]{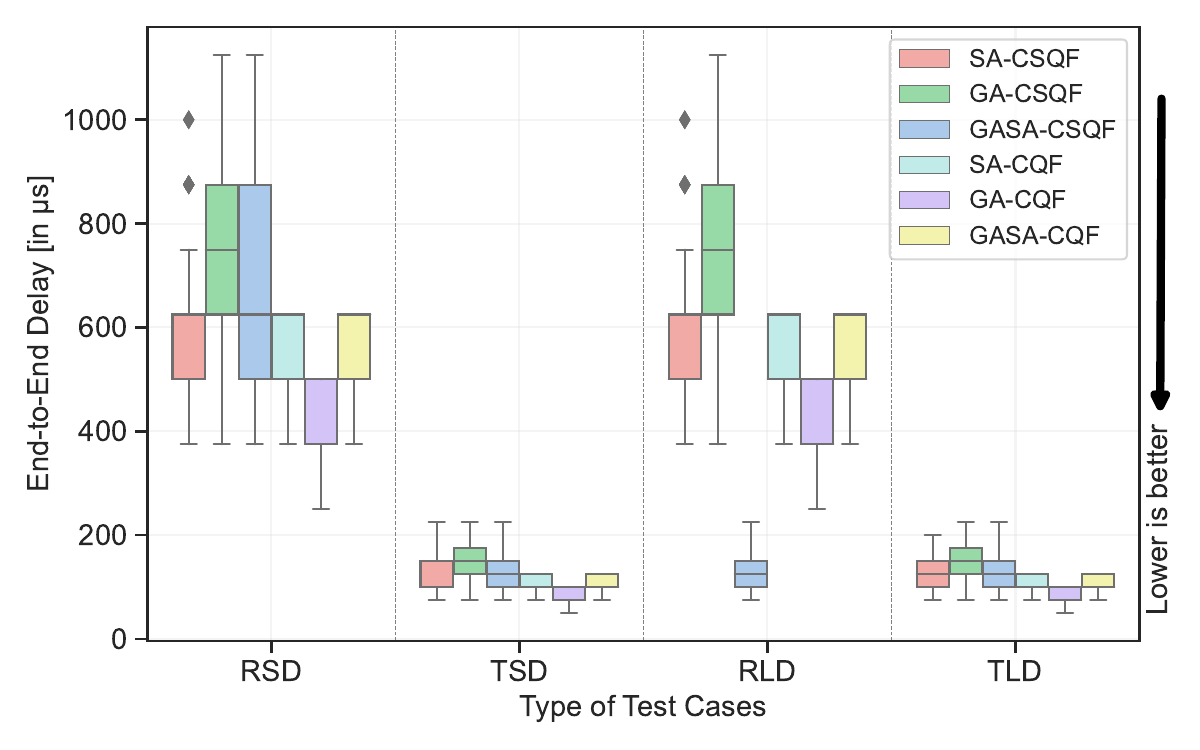}
    \subcaption{WCD} 
\label{fig:cqf_csqf_RRG_boxplot_delay}
    \end{minipage}
    \caption{Comparison of CQF and CSQF for RRG topology (Fig.~\ref{fig:rrg}) with 100 Mbps BW. For the Tight test case, $T = 25\mu s$ for both CQF and CSQF, while for the Relaxed test case, $T = 125\mu s$.} 
    \label{fig:cqf_csqf_RRG}
    \vspace{-0.1cm}
\end{figure*}

\begin{figure*}[!t]
    \centering
    \begin{minipage}[b]{0.48\textwidth}
    \centering
    \includegraphics[scale=0.42, trim={0.4cm 0.4cm 0.5cm 0.4cm}, clip]{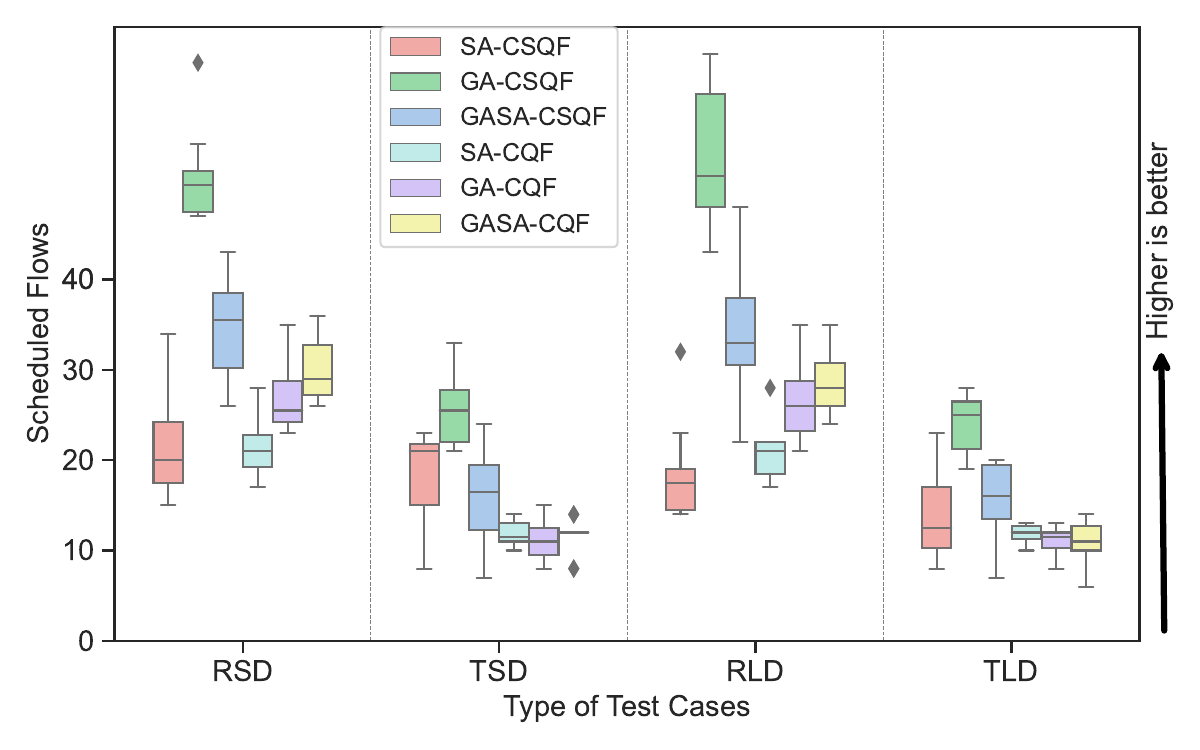}
    \subcaption{Schedulability}
    \label{fig:box_plot_cqf_csqf_schedule_bag}
    \end{minipage}
    \begin{minipage}[b]{0.48\textwidth}
    \centering
    \includegraphics[scale=0.42, trim={0.4cm 0.4cm 0.5cm 0.4cm}, clip]{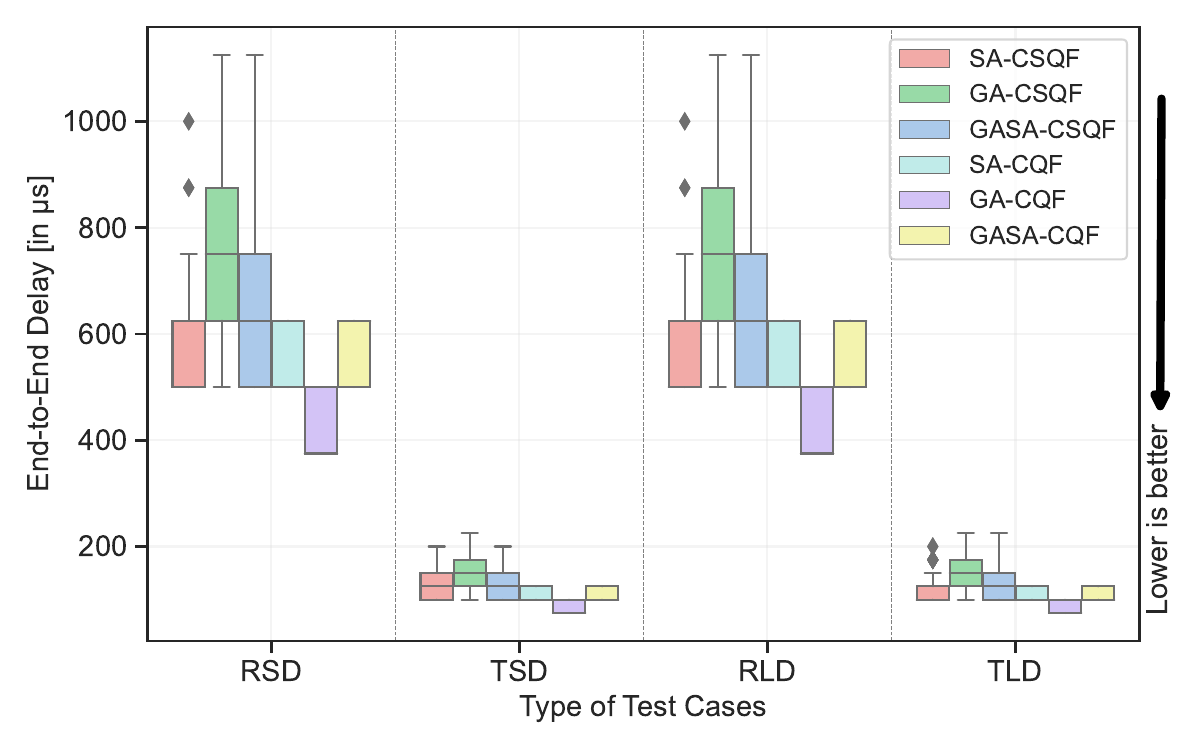}
    \subcaption{WCD} 
\label{fig:cqf_csqf_BAG_boxplot_delay}
    \end{minipage}
    \caption{Comparison of CQF and CSQF for BAG Topology (Fig.~\ref{fig:bag}) with 100 Mbps BW. For the Tight test case, $T = 25\mu s$ for for both CQF and CSQF, while for the Relaxed test case, $T = 125\mu s$.}
    \label{fig:cqf_csqf_BAG}
\end{figure*}

\subsection{Genetic Algorithm with Simulated Annealing (GASA)}
\label{sec:gasa}
Because of the run-time advantage of SA, we chose to incorporate the SA search into the GA algorithm (GASA) to increase the number of explored solutions in a short time span and, at the same time, benefit from the population search of the GA algorithm. In GASA, we replace the mutation operator of GA with the SA mutation, which is an SA search, and the pairing operator with the SA recombination (SA-R). With this pairing operator, we modify the parent selection part, elitism, by allowing the weaker solutions to be accepted into the next generation of the population with a certain probability that depends on the artificial temperature of SA. In GA, the mutation is a random process (refer to Algorithm~\ref{alg:ga}), however, the challenge is getting stuck at local optima. To overcome this, GASA provides us with the advantages of both GA and SA by providing us with localized SA search. Therefore, instead of random mutation, we use the SA search operation to perform the mutation in GASA (Alg.~\ref{alg:pseudo_crossover}, line 10). With GASA, due to the SA search, we avoid getting stuck at the local optima.

\begin{figure*}[!t]
    \centering
    \begin{minipage}[b]{0.48\textwidth}
    \centering
    \includegraphics[scale=0.42, trim={0.4cm 0.4cm 0.5cm 0.4cm}, clip]{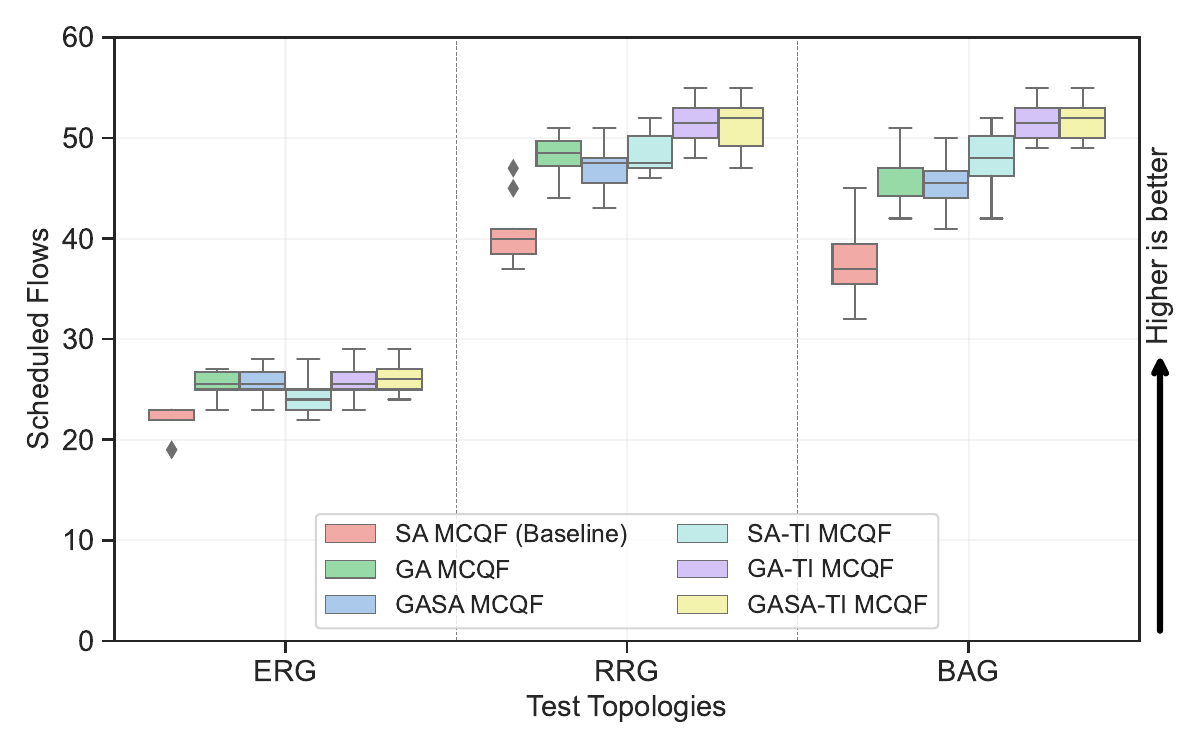}
    \subcaption{Multi-CQF Schedulability} 
    \label{fig:mcqf_box_erg_rrg_bag}
    \end{minipage}
    \begin{minipage}[b]{0.48\textwidth}
    \centering  
    \includegraphics[scale=0.42, trim={0.4cm 0.4cm 0.5cm 0.4cm}, clip]{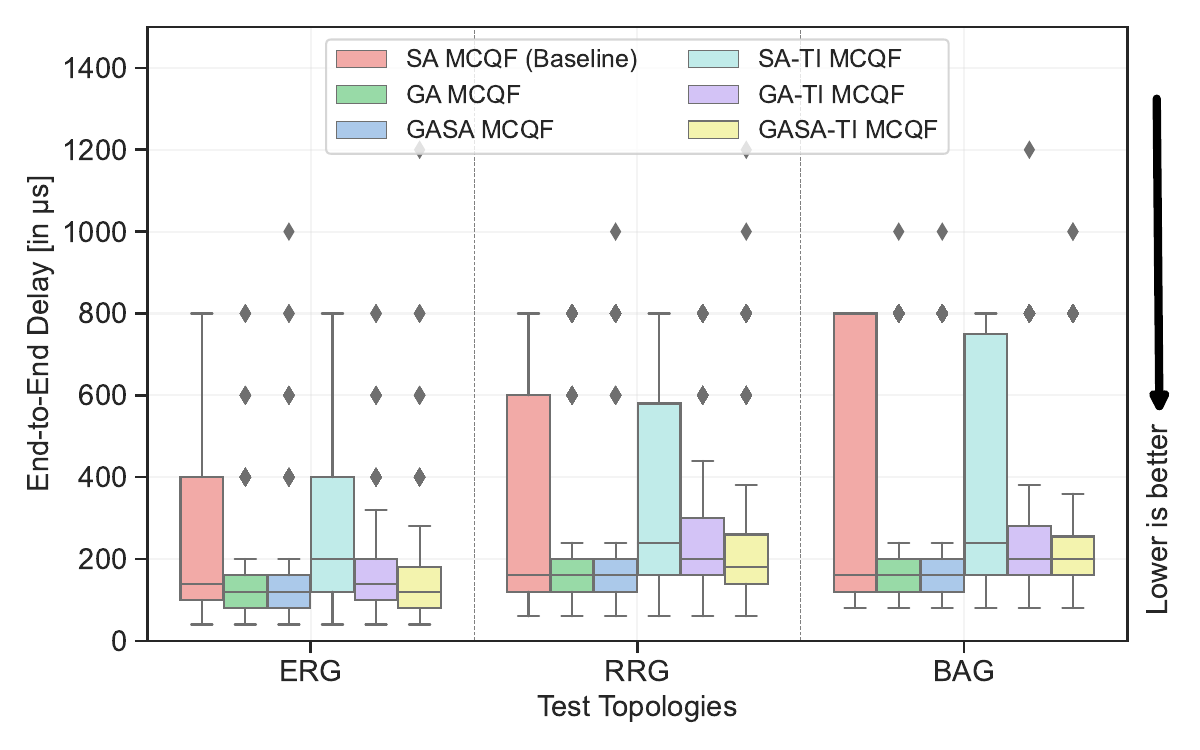}
    \subcaption{Multi-CQF WCD} 
 \label{fig:mcqf_box_erg_rrg_bag_delay}
    \end{minipage}
    \caption{Comparison of algorithms for Multi-CQF across different topologies (ERG, RRG, and BAG) with 100 Mbps BW and cycle of (125, 250, 500) $\mu$s for RLD TC.}
    \label{fig:mcqf_box_erg_rrg_bag_all}
    \vspace{-0.2cm}
\end{figure*}

\vspace{-0.1cm}
\section{Evaluation and Discussion}
\label{sec:results}
We implemented the algorithms on a Windows laptop with an Intel\textregistered\ Processor Core\texttrademark\ i7-10610U CPU running at 1.80 GHz, and 32 GB RAM. The proposed algorithms generate configurations for CQF, CSQF, and Multi-CQF. To compare our work with the related work, we implemented the SA algorithm following the work in \cite{mcqf_paul}. The proposed models terminate if the solution does not improve after a predefined number of iterations. After tuning, we set $\alpha$ to 2 and $\beta$ to 3 for the evaluation. We tested our model with different types of Test Cases (TCs): 

\begin{enumerate}
    \item \textbf{Relaxed Small Deadline (RSD):} In this set of TCs, the flows have periodicity of 1000, 2500, 5000, or 10000 $\mu$s, the size is between 55-1500 Bytes, and the deadline of the flows $<$ the periodicity of the flows.
    \item \textbf{Tight Small Deadline (TSD):} In this TC, the periodicity of the flows are 100, 500, or 1000 $\mu$s, the size is between 55-200 Bytes, and the deadline of the flow $<$ the periodicity of the flow.
    \item \textbf{Relaxed Large Deadline (RLD):} In this TC, the periodicity of the flow is 1000, 2500, 5000, or 10000 $\mu$s, the size is between 55-1500 Bytes, and the deadline of the flow is $=$ the periodicity of the flow.
    \item \textbf{Tight Large Deadline (TLD):} In this TC, the periodicity is 100, 500, or 1000 $\mu$s, the size is between 55-200 Bytes, and the deadline of the flow $=$ the periodicity of the flow.
\end{enumerate}

In this paper, we consider only one traffic type in the network (TT traffic). Furthermore, all the TT flows are non-preemptable and there is no frame preemption~\cite{rubi_noms} enabled in the network. Each topology (refer to Fig.~\ref{fig:erg},~\ref{fig:rrg},~\ref{fig:bag},~\ref{fig:ring}) is tested with ten different TCs. We compare the performance of our algorithms using the following metrics: 

\begin{enumerate}
    \item \textbf{Schedulability of TT flow:} This metric represents the total scheduled flows by the algorithm. \textbf{Higher is better.}
    \item \textbf{Computation cost of the algorithm:} The total run time of the algorithm. \textbf{Lower is better.}
    \item \textbf{Convergence of the algorithm:} This metric shows the objective function convergence time for the algorithms. \textbf{Faster is better.}
\end{enumerate}

Our algorithms are specifically designed for Multi-CQF. However, by configuring them with a single instance of CQF or CSQF, we are able to evaluate their behavior in CQF and CSQF networks as well. We would like to highlight that Multi-CQF does not necessarily schedule a larger number of flows than CQF or CSQF across all TCs. The key advantage of Multi-CQF lies in its ability to support traffic with diverse timing requirements, which is a limitation in current CQF and CSQF. In all experiments, unless otherwise specified, we use the DBM mechanism for flow-to-\texttt{QG} mapping and apply a fixed BW distribution of 40\% to \texttt{QG}1, 30\% to \texttt{QG}2, and 20\% to \texttt{QG}3. The remaining 10\% is left for the BE traffic type. This BW configuration is chosen to match the BW distribution used in \cite{mcqf_paul}, allowing for consistent and fair comparison with the baseline. We also perform additional experiments to evaluate how different flow-to-\texttt{QG} mappings and cycle combinations affect the overall schedulability of Multi-CQF. 

\begin{figure*}[!t]
\centering
\includegraphics[scale=0.62, trim={0.3cm 0.4cm 0.3cm 0.3cm}, clip]{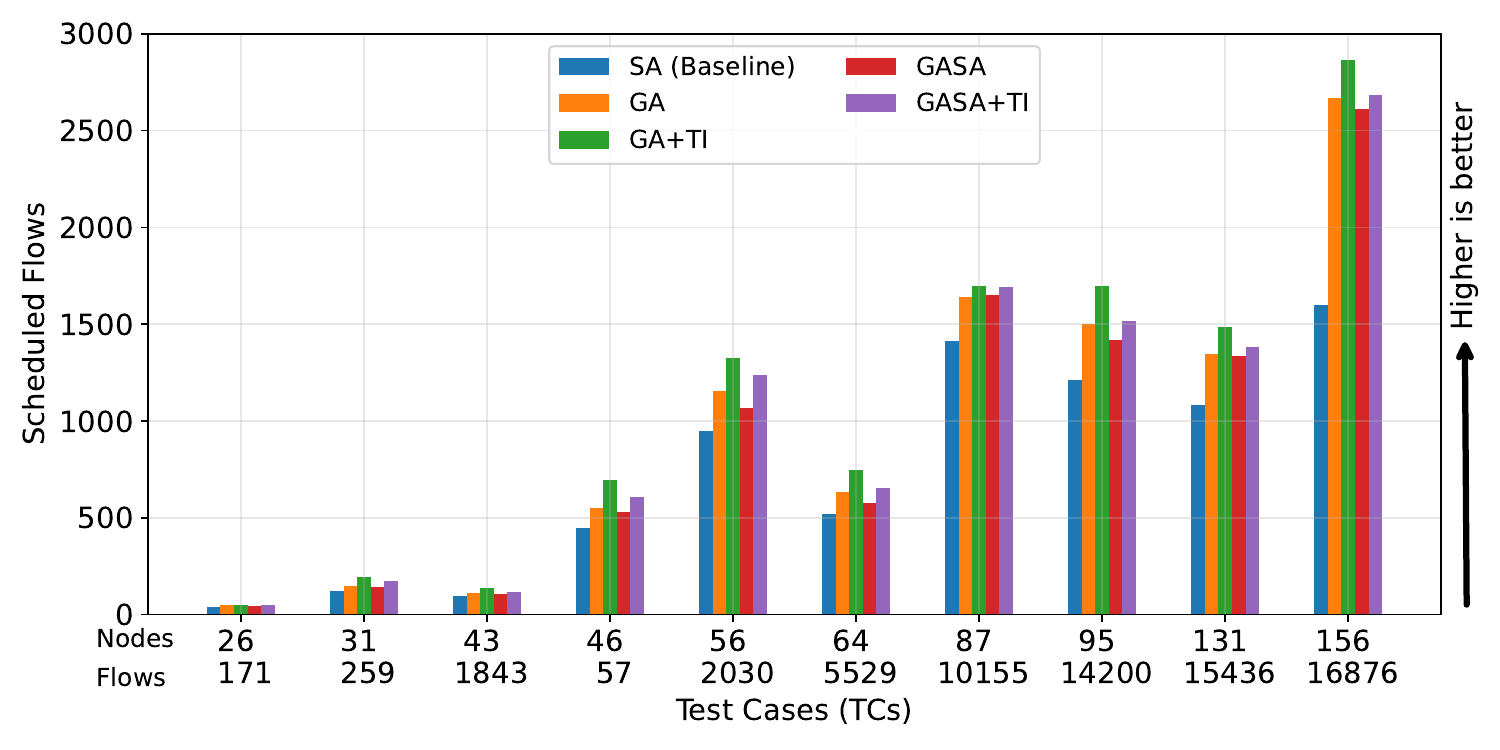}
\caption{Multi-CQF schedulability across different algorithms for the Ring topology with 1 Gbps BW and a cycle combination of (25, 50, 100) $\mu$s}.
\label{fig:mcqf_bar_plot_flows}
\vspace{-0.1cm}
\end{figure*}

\begin{figure*}[!t]
    \centering
    \includegraphics[scale=0.46, trim={0.4cm 0.4cm 3.5cm 0.3cm}, clip]{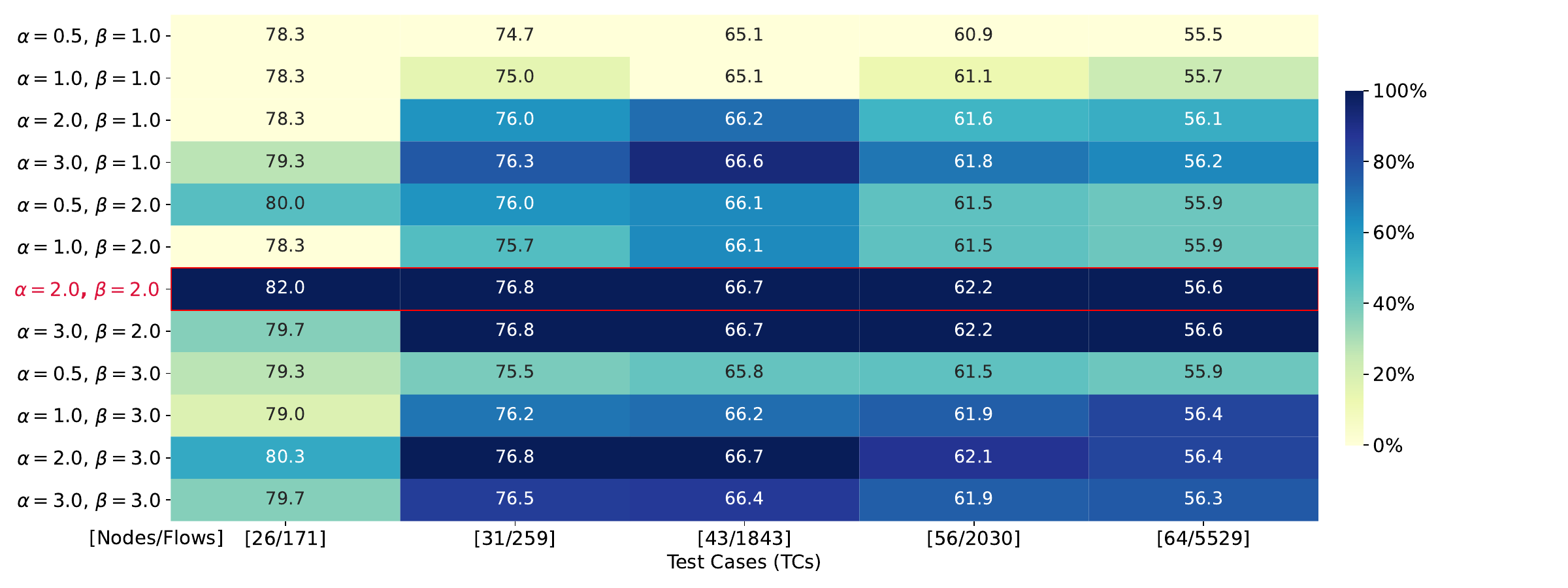}
    \caption{Comparison of different values of $\alpha$ and $\beta$ for Multi-CQF across various TCs on the Ring topology. The values indicate the schedulability percentage of flows for each $\alpha$ and $\beta$ combination in the given TC, using the GASA algorithm with 1 Gbps BW and cycle of (25, 50, 100) $\mu$s. As it is seen from the result, the schedulability is highest for $\alpha$ = 2, and $\beta$ = 2.}
    \label{fig:alpha_beta_mcqf}
    \vspace{-0.1cm}
\end{figure*} 

\subsection{CQF and CSQF}
\label{sub:cqf} 
First, we evaluate CQF and CSQF using different algorithms on ERG (Fig.~\ref{fig:erg}), RRG (Fig.~\ref{fig:rrg}), and BAG (Fig.~\ref{fig:bag}) topologies with 100 Mbps BW. All algorithms use the same TC, network parameters, and cycle. The x-axis in Fig.~\ref{fig:cqf_csqf_ER_boxplot_schedule}, \ref{fig:box_plot_cqf_csqf_RRG_schedule}, and \ref{fig:box_plot_cqf_csqf_schedule_bag} represents the TC type (RSD, TSD, RLD, and TLD), while the y-axis shows the number of scheduled flows. For each topology (ERG, RRG, and BAG) and TC type, we run ten different TCs to ensure robustness and consistency in the evaluation. Each algorithm is evaluated on all ten TCs for each TC type. The box plots illustrate the distribution of the number of scheduled flows for each algorithm, with separate plots presented for each TC type to enable comparison across different flow distribution patterns. As shown in Fig.~\ref{fig:cqf_csqf_ER_boxplot_schedule}, \ref{fig:box_plot_cqf_csqf_RRG_schedule}, and \ref{fig:box_plot_cqf_csqf_schedule_bag}, CSQF schedules more flows with all the three algorithms (SA, GA, and GASA) due to the \emph{tolerating queue}, however, the delay is also more prominent in CSQF. The TT flows mapped to the third queue always have to wait for an additional cycle attributing to the increased delay. Our proposed GA and GASA algorithms outperform the baseline SA by scheduling significantly larger number of flows (as highlighted in Fig.~\ref{fig:cqf_csqf_ER_boxplot_schedule}, \ref{fig:box_plot_cqf_csqf_RRG_schedule}, and \ref{fig:box_plot_cqf_csqf_schedule_bag}) for all the three different topologies. The WCDs (refer Fig.~\ref{fig:cqf_csqf_ER_boxplot_delay}, \ref{fig:cqf_csqf_RRG_boxplot_delay}, and \ref{fig:cqf_csqf_BAG_boxplot_delay}) clearly show that CSQF leads to larger end-to-end delay compared to CQF. The results show that CSQF schedules a larger number of flows at the cost of increased end-to-end delay. Furthermore, CSQF does not always outperform CQF, as the tolerating queue is particularly useful when the propagation delay and synchronization error is not bounded (small and large) in the network. We showed similar behaviors using an OMNeT++ simulation-based approach in our previous work in \cite{rubi_ccnc} where CSQF did not show its dominance all the time over CQF. Fig.~\ref{fig:cqf_csqf_ER_boxplot_delay}, \ref{fig:cqf_csqf_RRG_boxplot_delay}, and \ref{fig:cqf_csqf_BAG_boxplot_delay}, show that GA and GASA CSQF have larger delays compared to SA CSQF. This is because the GA and GASA algorithms aim to increase the schedulability at the cost of higher WCD by using the tolerating queue more often. 

\subsection{Evaluation of Multi-CQF}
\label{sub:eval_mcqf}
For evaluating Multi-CQF, we test four different topologies - ERG, RRG, BAG, and Ring (refer to Fig.~\ref{fig:erg_rrg_bag_ring}). Fig.~\ref{fig:mcqf_box_erg_rrg_bag} compares the schedulability performance of the proposed algorithms with the baseline method across ERG, RRG, BAG topologies. In Fig.~\ref{fig:mcqf_box_erg_rrg_bag_all}, the x-axis represents the topology type. For all three topologies (ERG, RRG, and BAG), we use the RLD TC type with cycle of (125, 250, 500) $\mu$s. All algorithms use the same TC, network parameters and cycle configuration.
GA Multi-CQF and GASA Multi-CQF schedule more flows than the baseline SA method. With the introduction of TI, the number of flows scheduled further increases as shown in Fig.~\ref{fig:mcqf_box_erg_rrg_bag}. However, as TI deliberately delays flows in the sending node, the WCD increases (see Fig.~\ref{fig:mcqf_box_erg_rrg_bag_delay}). Despite introducing TI, both GA and GASA achieve a similar end-to-end delay for the Multi-CQF network as SA. We further evaluate the performance of Multi-CQF in varying network sizes using the Ring topology (Fig.~\ref{fig:ring}). As in previous experiments, all algorithms use the same TC, network parameters, and cycle configuration. We increased the network load by running the experiment with more than 16000 flows for 1Gbps BW. In Fig.\ref{fig:mcqf_bar_plot_flows}, the x-axis represents the different TCs, along with the corresponding total number of nodes and flows in the network. The cycles are set to (25, 50, 100) $\mu$s. The y-axis shows the total number of scheduled flows for the Ring topology across different algorithms. Across all network load conditions, GA+TI and GASA+TI outperform the baseline SA. The introduction of TI clearly improves on the total scheduled flows by delaying the flows in the source node. 

\subsection{Hyperparameter Tuning of $\alpha$ and $\beta$}
\label{sub:alpha_beta}
Fig.~\ref{fig:alpha_beta_mcqf} presents the hyperparameter tuning results for our model under different values of $\alpha$ and $\beta$. We perform this tuning for Multi-CQF across different TCs in the Ring topology, using a 1 Gbps link speed and cycle of (25, 50, 100) $\mu$s. The tuning is carried out using the GASA algorithm, as it provides a better trade-off between schedulability and runtime compared to the GA algorithm (discussed below in Section~\ref{sub:convergence}). By varying $\alpha$ and $\beta$, we evaluate how the trade-off between their values affects overall schedulability in the Multi-CQF network, enabling us to select appropriate parameters for improved performance. As seen in Fig.~\ref{fig:alpha_beta_mcqf}, schedulability is highest across different test cases when $\alpha = 2$ and $\beta = 2$. This region is highlighted with a red box in Fig.~\ref{fig:alpha_beta_mcqf}.

\begin{figure*}[!t]
    \centering
    \begin{minipage}[b]{0.48\textwidth}
    \centering 
    \includegraphics[scale=0.4, trim={0cm 0.4cm 0.5cm 0.2cm}, clip]{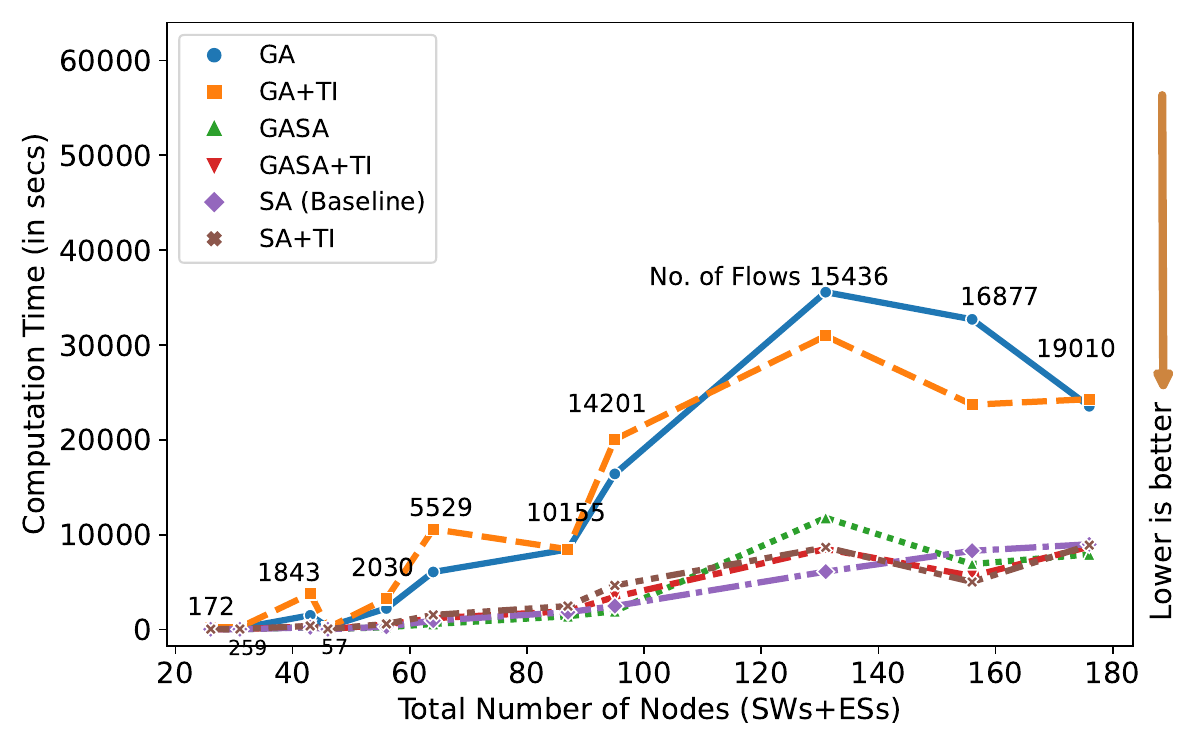}
    \subcaption{Time Complexity} 
    \label{fig:time_computation}
    \end{minipage}
    \begin{minipage}[b]{0.48\textwidth}
    \centering
    \includegraphics[scale=0.4, trim={0cm 0.4cm 0.5cm 0.2cm}, clip]{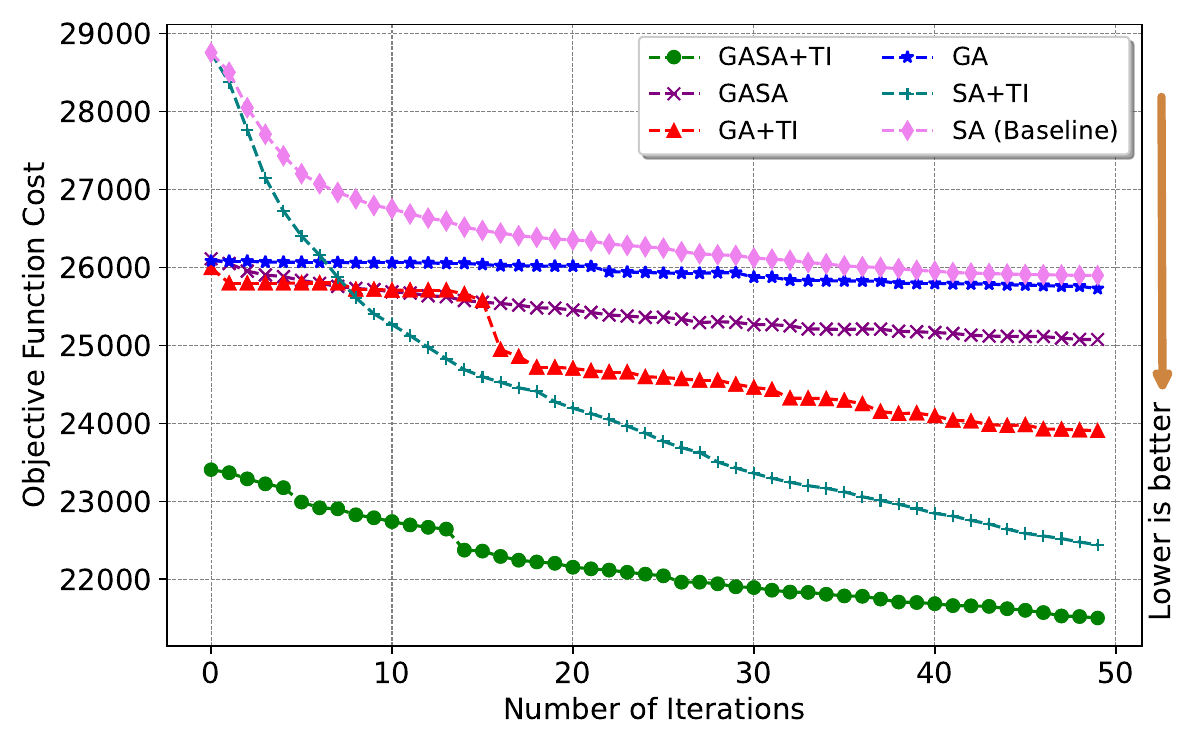}
    \subcaption{Objective Convergence} 
   \label{fig:objective_convergence}
    \end{minipage}
    \caption{Comparison of different Multi-CQF algorithms for Ring topology with 1 Gbps link speed: (a) Time, and (b) Objective.}
    \label{fig:obj_comp}
    \vspace{-0.2cm}
\end{figure*}

\begin{figure*}[!t]
    \centering
    \begin{minipage}[b]{0.47\textwidth}
    \centering 
    \includegraphics[scale=0.52, trim={0.4cm 0.2cm 0.2cm 0.2cm}, clip]{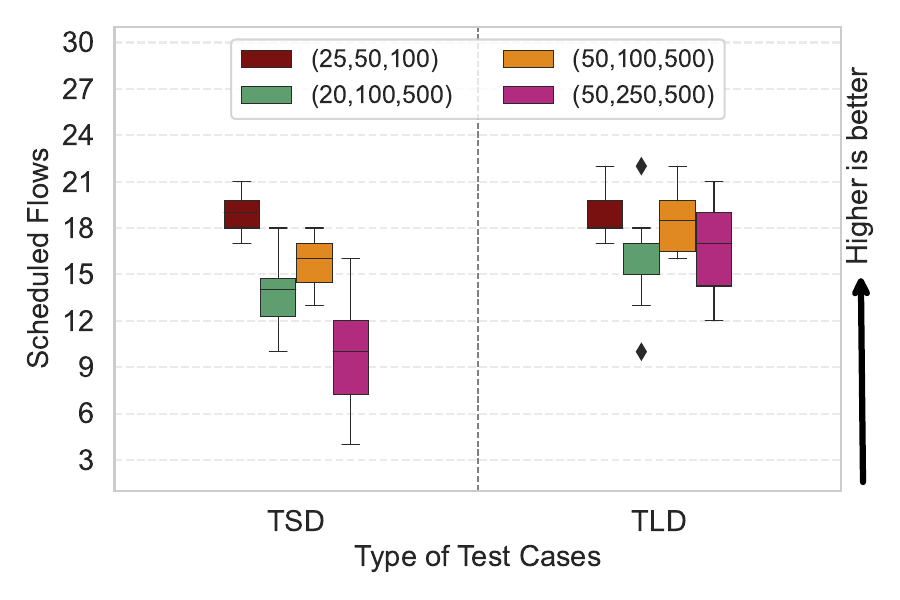}
    \subcaption{GA: Tight TC}
    \label{fig:erg_tight_cycle_ga}
    \end{minipage}
    \hfill
    \begin{minipage}[b]{0.47\textwidth}
    \centering 
    \includegraphics[scale=0.52, trim={0.4cm 0.2cm 0.2cm 0.2cm}, clip]{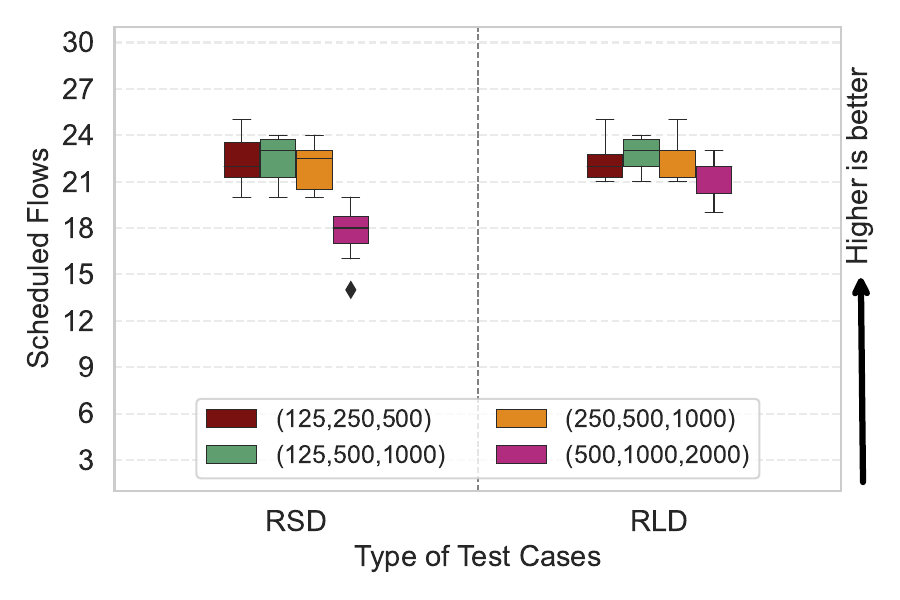}
    \subcaption{GA: Relaxed TC}
    \label{fig:erg_relaxed_cycle_ga}
    \end{minipage}
    \caption{Impact of different cycle combinations on Multi-CQF network performance for the ERG topology with 100 Mbps BW using the GA algorithm: (a) Tight TC (TSD, TLD), (b) Relaxed TC (RSD, RLD).}
\label{fig:erg_cycle_related_change_ga}
    \vspace{-0.2cm}
\end{figure*}

\begin{figure*}[!t]
    \centering
    \begin{minipage}[b]{0.47\textwidth}
    \centering 
    \includegraphics[scale=0.52, trim={0.4cm 0.2cm 0.2cm 0.2cm}, clip]{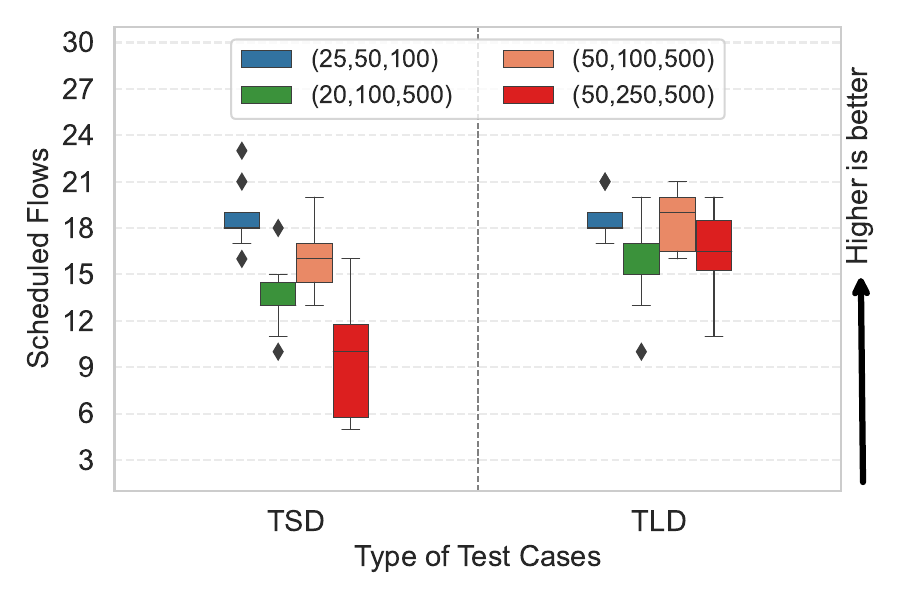}
    \subcaption{GASA: Tight TC}
    \label{fig:erg_tight_cycle_gasa}
    \end{minipage}
    \hfill
    \begin{minipage}[b]{0.47\textwidth}
    \centering 
    \includegraphics[scale=0.52, trim={0.4cm 0.2cm 0.2cm 0.2cm}, clip]{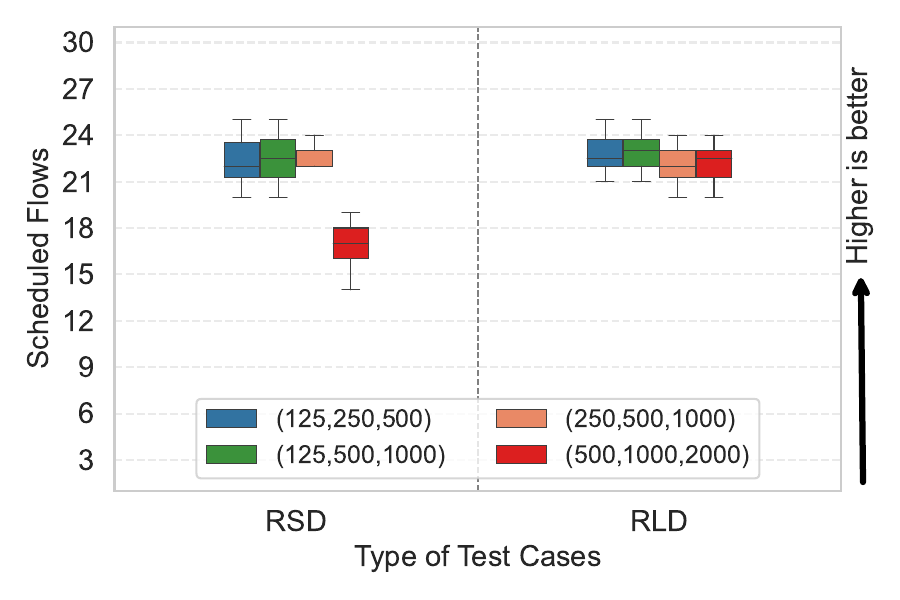}
    \subcaption{GASA: Relaxed TC}
    \label{fig:erg_relax_cycle_gasa}
    \end{minipage}
    \caption{Impact of different cycle combinations on Multi-CQF network performance for the ERG topology with 100 Mbps BW using the GASA algorithm: (a) Tight TC (TSD, TLD), (b) Relaxed TC (RSD, RLD).}
\label{fig:erg_cycle_related_change_gasa}
    \vspace{-0.2cm}
\end{figure*}

\vspace{-0.2cm}
\subsection{Time Complexity and Convergence}
\label{sub:convergence}
Time complexity is a key performance indicator for any algorithm. Given the pressing need for fast and scalable algorithms, studying time complexity becomes crucial. Therefore, we evaluated the time complexity and the convergence of the proposed algorithms shown in Fig.~\ref{fig:obj_comp}. SA has the least time complexity, while GA has the highest. However, GASA can provide reasonable time complexity, which is at the same level as SA but considerably smaller than GA (refer Fig.~\ref{fig:time_computation}). On the other hand, the objective function convergence (Fig.~\ref{fig:objective_convergence}) shows that GASA and GASA+TI converge the fastest. The convergence shows how fast our model converges to the optimal or near optimal solution. SA (baseline), on the other hand, is the slowest in terms of convergence. 

\begin{figure*}[!t]
    \centering
    \begin{minipage}[b]{0.44\textwidth}
    \centering 
    \includegraphics[scale=0.52, trim={0.4cm 0.2cm 0.4cm 0.2cm}, clip]{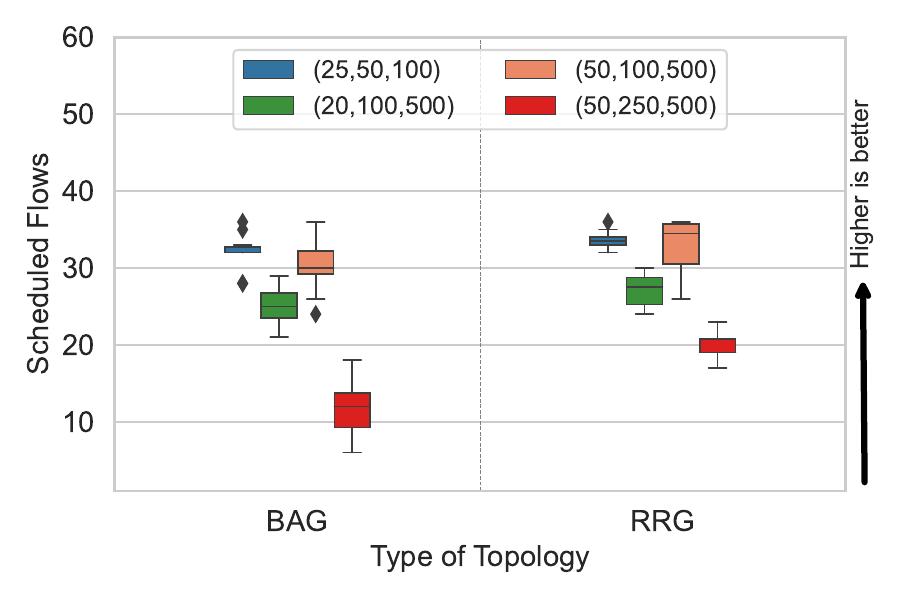}
    \subcaption{TSD TC}
    \label{fig:cycle_bag_rrg_tsd}
    \end{minipage}
    \hfill
    \begin{minipage}[b]{0.44\textwidth}
    \centering 
    \includegraphics[scale=0.52, trim={0.4cm 0.2cm 0.4cm 0.2cm}, clip]{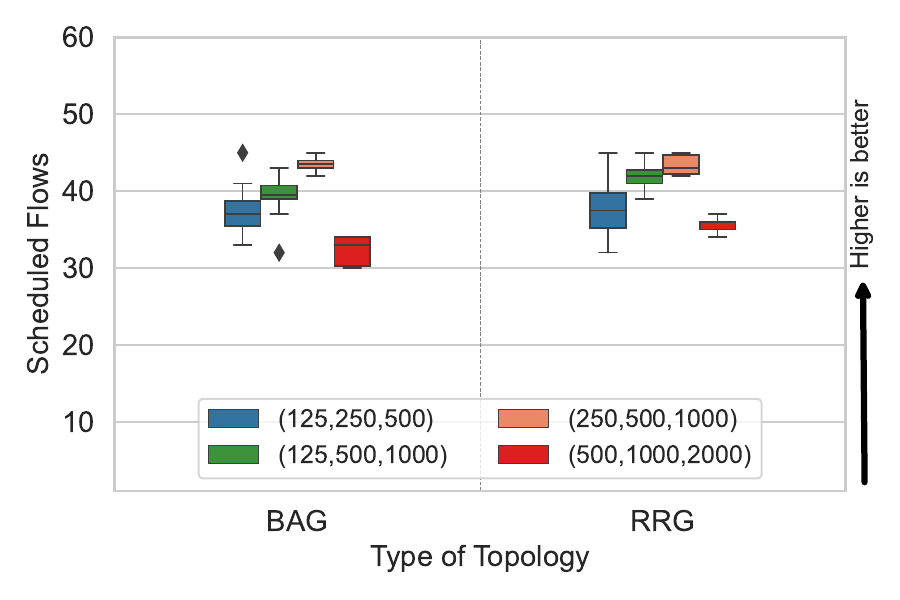}
    \subcaption{RSD TC}
     \label{fig:cycle_bag_rrg_rsd}
    \end{minipage}
    \caption{Comparison of different cycle combinations on BAG and RRG topologies with 100 Mbps BW using GASA algorithm: (a) TSD TC, and (b) RSD TC.}
    \label{fig:cycle_bag_rrg}
    \vspace{-0.2cm}
\end{figure*}

\begin{figure*}[!t]
    \centering
    \begin{minipage}[b]{0.44\textwidth}
    \centering 
    \includegraphics[scale=0.44, trim={0.4cm 0.2cm 0.4cm 0.2cm}, clip]{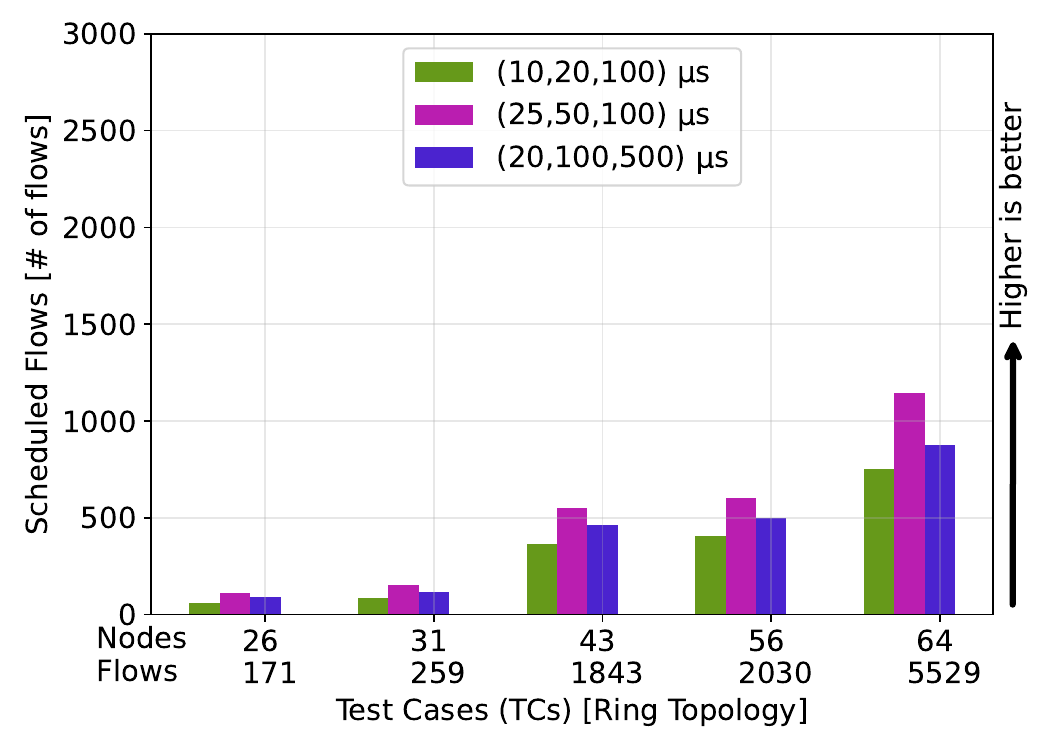}
    \subcaption{GA Algorithm}
    \label{fig:cycle_ring_ga}
    \end{minipage}
    \hfill
    \begin{minipage}[b]{0.44\textwidth}
    \centering 
    \includegraphics[scale=0.44, trim={0.4cm 0.2cm 0.4cm 0.2cm}, clip]{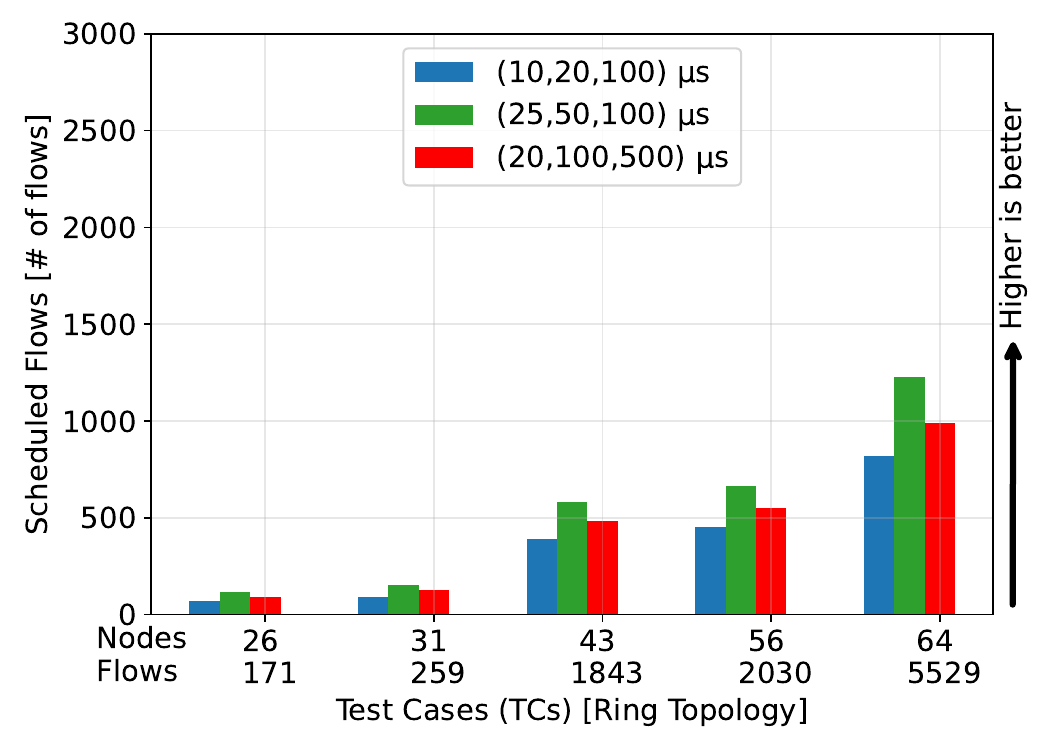}
    \subcaption{GASA Algorithm}
     \label{fig:cycle_ring_gasa}
    \end{minipage}
    \caption{Impact of different cycle combinations on overall Multi-CQF network performance for the Ring topology with 1 Gbps BW: (a) using the GA algorithm, and (b) using the GASA algorithm.}   \label{fig:cycle_compare_mcqf_ring}
    \vspace{-0.2cm}
\end{figure*}

\subsection{Impact of Cycle on Multi-CQF}
\label{sub:cycle_multi_cqf}
To evaluate the performance of Multi-CQF under different cycle combinations, we perform a schedulability analysis across ERG, RRG, BAG and Ring topologies. First, we start with the ERG topology where we compare the performance of different cycle combination using both GA and GASA algorithm across different TCs (TSD, TLD, RSD, and RLD). Fig.~\ref{fig:erg_cycle_related_change_ga} presents the schedulability analysis using the GA algorithm and Fig.~\ref{fig:erg_cycle_related_change_gasa} presents the schedulability analysis using the GASA algorithm. Fig.~\ref{fig:erg_tight_cycle_ga} shows the results for TSD and TLD TCs, while Fig.~\ref{fig:erg_relaxed_cycle_ga} shows the results for RSD and RLD with GA algorithm. Furthermore, Fig.~\ref{fig:erg_tight_cycle_gasa} shows the results for TSD and TLD TCs, and Fig.~\ref{fig:erg_relax_cycle_gasa} shows the results for RSD and RLD with GASA algorithm. All TCs are evaluated under four different cycle combinations. The results clearly show that cycle significantly impacts overall schedulability. The impact is especially notable in TCs with tighter deadlines, such as TSD and RSD, where longer cycles can lead to missed deadlines and reduced schedulability. Based on these observations, we focus on the TSD and RSD TCs in the next step when evaluating the RRG and BAG topologies. Fig.~\ref{fig:cycle_bag_rrg} shows the schedulability results of BAG and RRG topologies, for TSD (refer to Fig.~\ref{fig:cycle_bag_rrg_tsd}) and RSD (refer to Fig.~\ref{fig:cycle_bag_rrg_rsd}) TC. Similar to the ERG topology, the schedulability in BAG and RRG also declines significantly under unsuitable cycle combinations. For ERG, RRG, and BAG, we ran ten different TCs for each box plot. 

We further assess the impact of different cycle combination on the Ring topology, analyzing schedulability using both GA and GASA algorithm across varying network sizes and traffic loads. Fig.~\ref{fig:cycle_ring_ga} highlights how cycle selection affects schedulability using GA algorithm, while Fig.~\ref{fig:cycle_ring_gasa} illustrates the effect on schedulability using GASA algorithm. As discussed, non-optimal cycle combinations reduce the schedulability in both GA and GASA algorithms. Fig.~\ref{fig:cycle_compare_mcqf_ring} emphasizes the importance of using an efficient or near-optimal cycle combination in Multi-CQF to maintain high schedulability. \textit{Furthermore, it is evident from the results that there is no one cycle combination which fits all TCs. Therefore, for every TC, we need to find the best cycle combination given the network topology and the flow parameters. In summary, the choice of cycle has a direct and significant impact on schedulability in Multi-CQF networks, highlighting the necessity of solving the cycle optimization problem.}

\begin{figure*}[!t]
    \centering
    \begin{minipage}[b]{0.48\textwidth}
    \centering 
    \includegraphics[scale=0.53, trim={0.4cm 0.2cm 0.4cm 0.2cm}, clip]{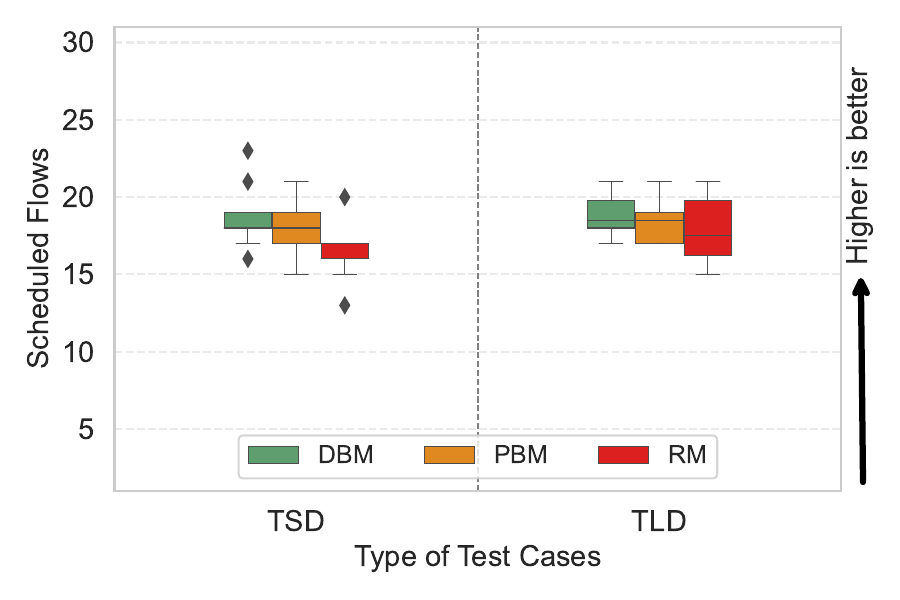}
    \subcaption{TSD and TLD TCs with a cycle of (25,50,100) $\mu$s.}
    \label{fig:sort_multi_cqf_tsd_tld}
    \end{minipage}
    \hfill
    \begin{minipage}[b]{0.48\textwidth}
    \centering
    \includegraphics[scale=0.53, trim={0.4cm 0.2cm 0.4cm 0.2cm}, clip]{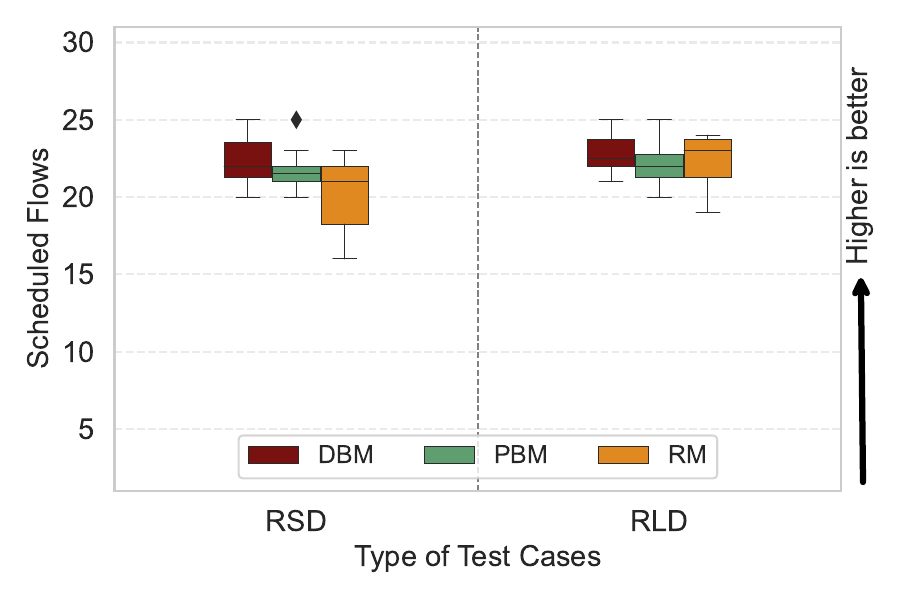}
    \subcaption{RSD and RLD TCs with a cycle of (125,250,500) $\mu$s.}
    \label{fig:sort_multi_cqf_rsd_rld}
    \end{minipage}
    \caption{Comparison of different flow sorting and flow-to-\texttt{QG} mapping mechanisms in Multi-CQF for the ERG topology using the GASA algorithm at 100 Mbps BW: (a) TSD and TLD TCs, and (b) RSD and RLD TCs.}
    \label{fig:sort_one}
   \vspace{-0.1cm}
\end{figure*}

\begin{figure*}[!t]
    \centering
    \begin{minipage}[b]{0.48\textwidth}
    \centering 
    \includegraphics[scale=0.53, trim={0.4cm 0.2cm 0.4cm 0.2cm}, clip]{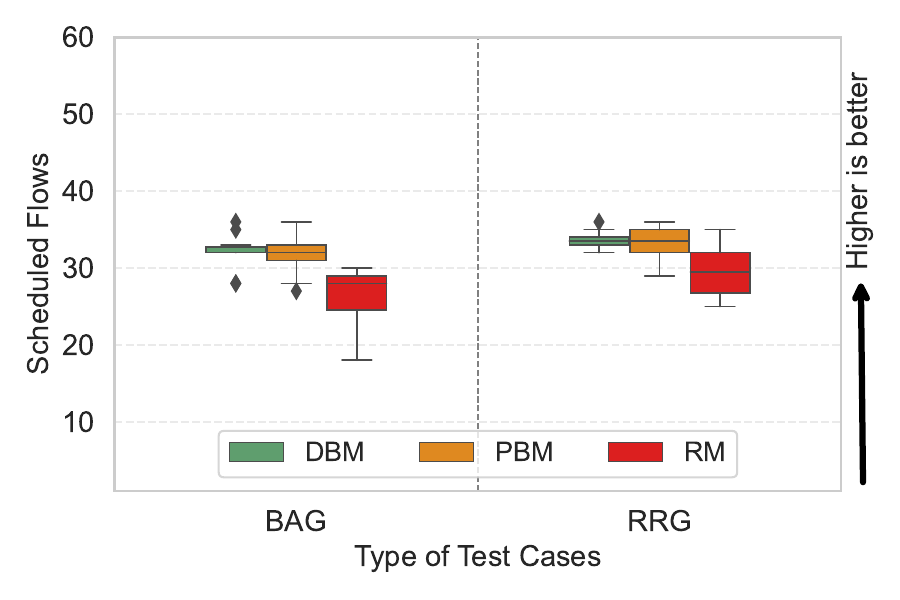}
    \subcaption{TSD TC with a cycle of (25,50,100) $\mu$s.}
    \label{fig:sort_bag_rrg_tsd}
    \end{minipage}
    \hfill
    \begin{minipage}[b]{0.48\textwidth}
    \centering
    \includegraphics[scale=0.53, trim={0.4cm 0.2cm 0.4cm 0.2cm}, clip]{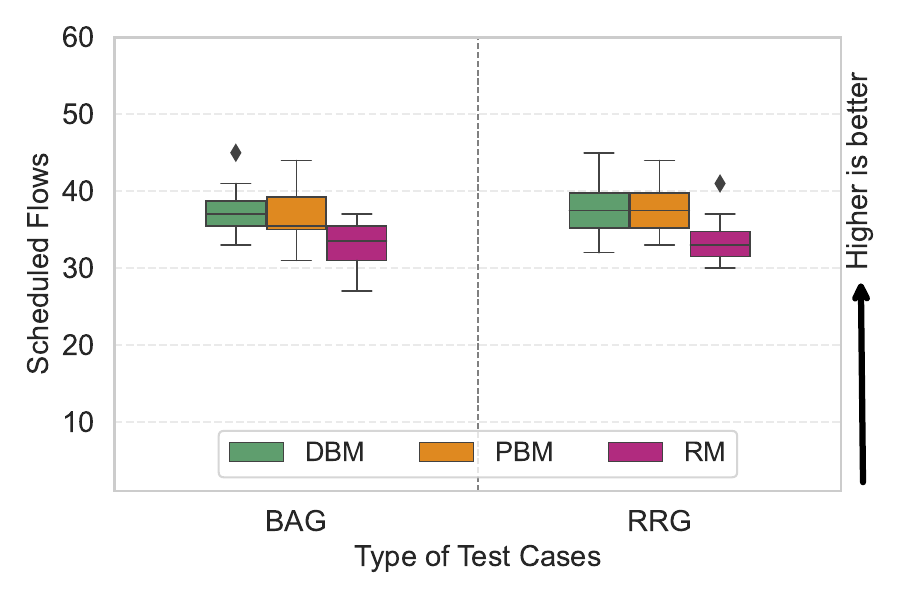}
    \subcaption{RSD TC with a cycle of (125,250,500) $\mu$s.}
    \label{fig:sort_bag_rrg_rsd}
    \end{minipage}
    \caption{Comparison of different flow-to-\texttt{QG} mapping mechanisms in Multi-CQF on BAG and RRG topologies with 100 Mbps BW using the GASA algorithm: (a) TSD TC, and (b) RSD TC.}
    \label{fig:sort_rrg_bag}
   \vspace{-0.1cm}
\end{figure*}

\subsection{Impact of flow sorting and flow-to-\texttt{QG} mapping}
\label{sub:flow_to_queue_group_multi_cqf}
We evaluate the impact of flow-to-\texttt{QG} mapping on overall schedulability by conducting a set of experiments across the ERG, RRG, and BAG topologies. We first sort the flows and then assign them to different \texttt{QG}s using three mapping strategies: DBM, PBM, and RM (details in Section~\ref{sub:flow_to_queue_group_multi_cqf}). In Fig.~\ref{fig:sort_multi_cqf_tsd_tld} and \ref{fig:sort_multi_cqf_rsd_rld}, we show the schedulability results for the ERG topology under tight and relaxed TCs. As seen in the figures, DBM achieves the highest schedulability across all TC types (TSD, TLD, RSD, and RLD). RM mapping performs the worst across all scenarios, underlining the need for a DSK-based and informed flow-to-\texttt{QG} mapping strategy. 

To extend our analysis, we study the impact of mapping strategies on the RRG and BAG topologies using the TSD and RSD TC, where deadlines are particularly small. In Fig.~\ref{fig:sort_bag_rrg_tsd} and \ref{fig:sort_bag_rrg_rsd}, we observe a similar pattern where DBM performs the best, while RM results in the lowest schedulability. As we have discussed throughout this paper, Multi-CQF is well suited for handling traffic with diverse timing requirements. Mapping flows to \texttt{QG}s based on their deadlines using DBM leads to significantly better schedulability. In future work, we plan to design an algorithm that can learn or compute an optimal flow-to-\texttt{QG} mapping.

\vspace{0.3cm}
\section{Conclusion}
This paper presents the problem formulation for Multi-Cyclic Queuing and Forwarding (Multi-CQF) and proposes two open-source algorithms - Genetic Algorithm (GA) and hybrid GA-Simulated Annealing (GASA). Given the NP-hard nature of the Multi-CQF problem, we analyze the time complexity and convergence of the proposed algorithms and compare them with the baseline SA implementation from literature. The core idea of our approach is to determine the cycle combination, leverage domain-specific knowledge (DSK) to map Time-Triggered (TT) flows into the corresponding Multi-CQF Queue Groups, and regulate Time Injection (TI) to control when TT flows are sent from the source node. Instead of iterating over thousands of flows and mapping them to the Queue Group and the respective queue one by one, we segregate the flows into Queue Groups, which eventually reduces the search space leading to a faster convergence of the algorithm. We effectively utilized TI to enhance overall schedulability. Additionally, we introduce constraints to find the proper cycle combination and optimize the time complexity of Multi-CQF. The results demonstrate that both GASA and GA outperform the baseline SA algorithm by scheduling more TT flows at the cost of higher end-to-end delay. GASA is also statistically significantly faster than GA, approaching the time complexity of SA. We also demonstrate that improper cycle choices severely degrade schedulability, supporting the need for efficient cycle selection. Moreover, we analyze the impact of flow-to-Queue-Group mapping strategies and find that Deadline-Based Mapping yields the best results across all topologies and test cases. In summary, our paper clarifies the working mechanism and highlights the advantages of implementing TI in Multi-CQF across all three algorithms. 

\section*{Acknowledgments}
We would like to thank Selver Sadikovic for his contributions on the initial parts of this work which shaped aspects of this paper.

\bibliographystyle{IEEEtran}

\bibliography{reference}

\newpage

\section{Biography Section}

\vspace{-33pt}
\begin{IEEEbiography}[{\includegraphics[width=1in,height=1.25in,clip,keepaspectratio]{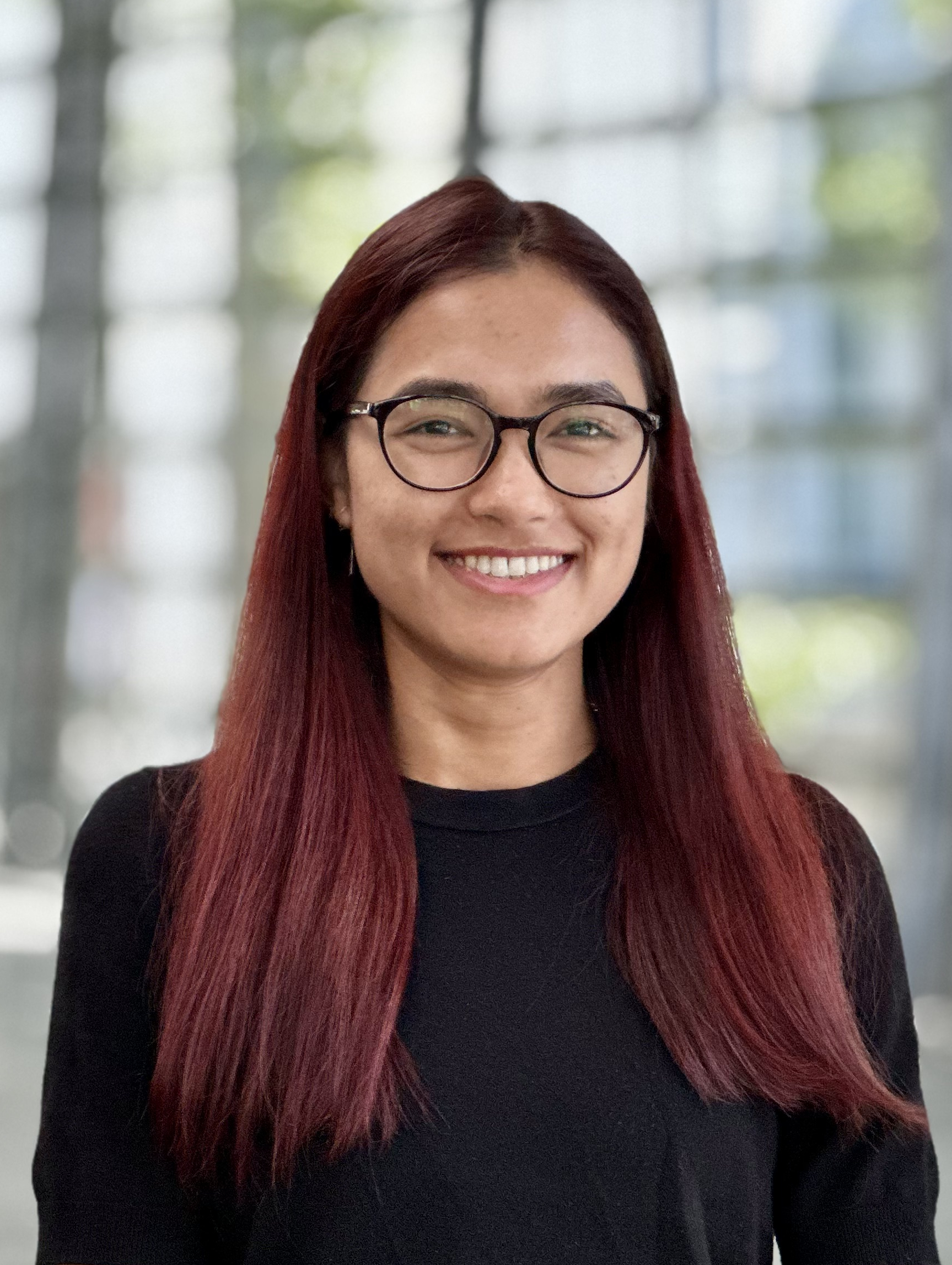}}]{Rubi Debnath} (Graduate Student Member, IEEE) received her Master of Science (M.Sc. TUM) degree in Communications Engineering from the Technical University of Munich (TUM), Munich, Germany, in 2020, graduating with distinction. After completing her master's, she joined the Associate Professorship of Embedded Systems and Internet of Things at the School of Computation, Information and Technology, Technical University of Munich, as a doctoral researcher and teaching associate. Her research interests include the design, optimization, and analysis of Time Sensitive Networking (TSN), 5G-TSN, and wireless TSN, using heuristic, metaheuristics, and machine learning models.
\end{IEEEbiography}

\vspace{-33pt}
\begin{IEEEbiography}[{\includegraphics[width=1in,height=1.25in,clip,keepaspectratio]{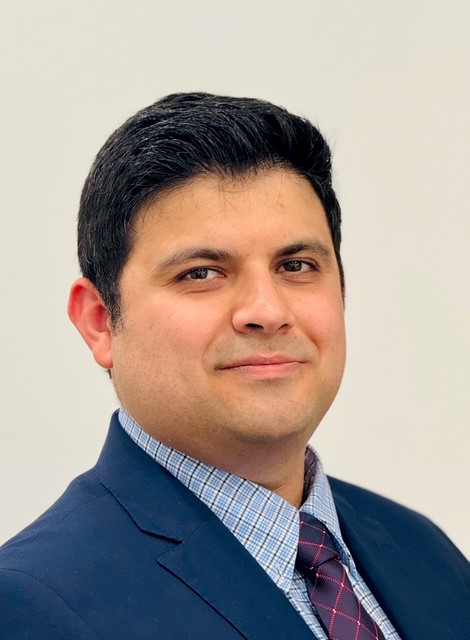}}]{Mohammadreza (Reza) Barzegaran} is a postdoctoral scholar at the Center for Pervasive Communications and Computing at the University of California, Irvine. He received his Ph.D. in Computer Science from the Technical University of Denmark (DTU) in 2021, where he was awarded the prestigious Marie Skłodowska-Curie Fellowship. Following his Ph.D., he held a postdoctoral position at DTU, where he served in a European training network program. Dr. Barzegaran holds an M.S. in Control Engineering from the University of Tehran and a B.S. in Aerospace Engineering from Amirkabir University of Technology. His research interests include wireless and wired sensor networks, real-time and safety-critical cyber-physical systems, and Fog/Edge computing platforms.
\end{IEEEbiography}
\vspace{-33pt}
\begin{IEEEbiography}[{\includegraphics[width=1in,height=1.25in,clip,keepaspectratio]{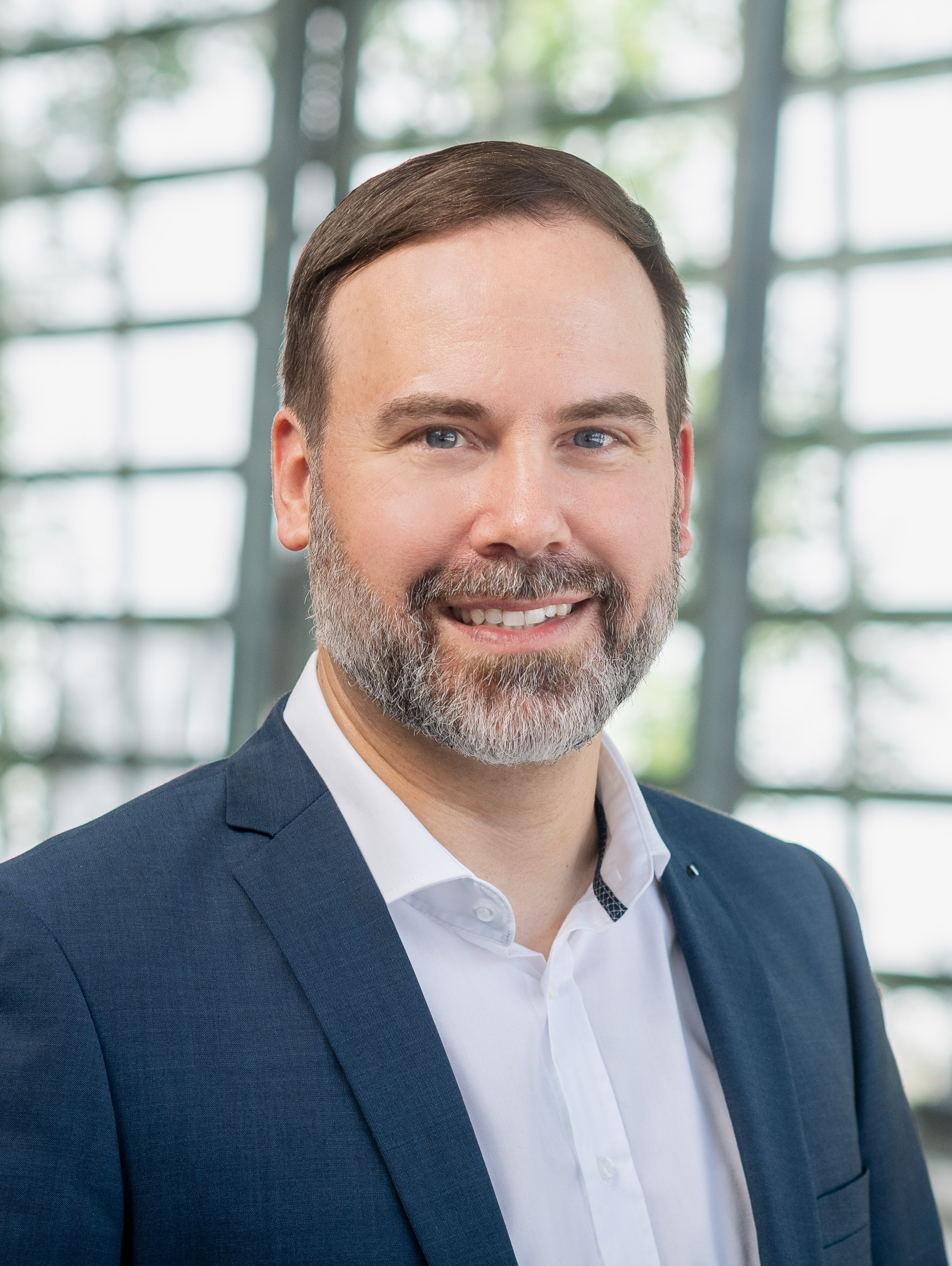}}]{Sebastian Steinhorst} (Senior Member, IEEE) received the M.Sc. (Dipl.-Inf.) and Ph.D. (Dr. phil. nat.) degrees in computer science from Goethe University, Frankfurt, Germany, in 2005 and 2011, respectively. He is currently an Associate Professor with the Department of Computer Engineering, Technical University of Munich (TUM), Germany, where he leads the Embedded Systems and Internet of Things Group. He was also a Co-Program PI with the Electrification Suite and Test Laboratory, Research Center TUMCREATE, Singapore. His research interests include design methodology and hardware/software architecture co-design of secure distributed embedded systems for use in IoT, Industry 4.0, and automotive.
\end{IEEEbiography}

\vfill

\end{document}